\documentclass[twoside]{article}

\usepackage{amsmath,amssymb}
\usepackage{url}
\usepackage{graphicx}

\newtheorem{theorem}{Theorem}
\newtheorem{proposition}[theorem]{Proposition}
\newtheorem{lemma}[theorem]{Lemma}

\newlength{\espacioproof}
\newenvironment{proof}[1][Proof]{\noindent \textbf{#1.} }{\ \rule{.4em}{.4em} \vspace{\espacioproof}}

\newcommand{\R}{\mathbb{R}}
\newcommand{\N}{\mathbb{N}}
\newcommand{\Z}{\mathbb{Z}}

\renewcommand{\a}{\alpha}
\renewcommand{\b}{\beta}

\newcommand{\e}{\varepsilon}
\newcommand{\g}{\gamma}
\newcommand{\G}{\Gamma}

\newcommand{\s}{\sigma}

\renewcommand{\t}{\theta}

\newcommand{\p}{\partial}

\renewcommand{\k}{\kappa}

\newcommand{\dd}{\, \mathrm{d}}

\DeclareMathOperator{\Card}{Card}

\DeclareMathOperator{\id}{id}
\newcommand{\mc}{\mathcal}

\title{Continuum limits of atomistic energies allowing smooth and sharp interfaces in 1D Elasticity}

\author{Carlos Mora-Corral}

\date{\small Mathematical Institute, University of Oxford. 24--29 St Giles'. Oxford OX1 3LB. UK \\
E-mail: \url{mora-cor@maths.ox.ac.uk}
}

\begin{document}
\maketitle

\begin{abstract}
In this paper we present two atomistic models for the energy of a one-dimensional elastic crystal.
We assume that the macroscopic displacement equals the microscopic one.
The energy of the first model is given by a two-body interaction potential, and we assume that the atoms follow a continuous and piecewise smooth macroscopic (continuum) deformation.
We calculate the first terms of the Taylor expansion (with respect to the parameter representing the interatomic distance) of the atomistic energy, and obtain that the coefficients of that Taylor expansion represent, respectively, an elastic energy, a sharp-interface energy, and a smooth-interface energy.
The second atomistic model is a variant of the first one, and its Taylor expansion predicts, in addition, a new term that accounts for the repulsion force between two sharp interfaces.
\end{abstract}

\section{Introduction}\label{se:intro}

The aim of this paper is to derive the continuum expression of the elastic energy and interfacial energy of a one-dimensional elastic crystal from an atomistic model.
The motivation of that analysis came from our desire to justify the continuum model proposed in Ball \& Mora-Corral \cite{BMC08}, according to which the same material can exhibit smooth and sharp interfaces.
In the one-dimensional case, that model is briefly described as follows.
The elastic solid is represented by the interval $(a,b)$, for some $a<b$.
An elastic deformation of the body $(a,b)$ is represented by an increasing, absolutely continuous map $u: (a,b) \to \R$ such that the function $u$ restricted to $(a,b) \setminus S$ is in the Sobolev space $W^{2,2}$, for some finite set $S$ depending on $u$.
Then, in \cite{BMC08} we postulated that equilibrium configurations are minimisers of the energy $I$ defined by
\begin{equation}\label{eq:I}
 I(u) := \int_a^b \left[ W(u'(x)) + \e^2 u''(x)^2 \right] \dd x + \k \Card (S_{u'}) ,
\end{equation}
where $S_{u'}$ is the set of discontinuity points of $u'$, and $\e,\k>0$ are two small parameters.

In \cite{BMC08} we saw that, when $W$ is assumed to have two wells, if $0<\k \ll \e \ll 1$ then the global minimisers $u$ of $I$ (subject to appropriate Dirichlet boundary conditions) satisfy $\Card S_{u'} =1$, and, in particular, they present sharp interfaces, whereas if $0<\e \ll \k \ll 1$, then the global minimisers satisfy $S_{u'} = \varnothing$ and present smooth interfaces.
This suggests that in the expression \eqref{eq:I}, one should replace $\k$ with a multiple of $\e$, since only when $\k$ is comparable to $\e$ can both smooth and sharp interfaces appear simultaneously.

In this paper, we try to derive the energy \eqref{eq:I} (once $\k$ has been substituted by a multiple of $\e$) from an atomistic one.
The main idea, which is based on the procedure of Blanc, Le Bris \& Lions \cite{BLL02}, is the following.
We assume that there is a continuum deformation $u : [a,b] \to \R$ such that, for every small $\e>0$, the atomistic (discrete) deformation is the restriction of $u$ to $[a,b] \cap \e\Z$, and the atomistic energy of that discrete deformation is given by
\begin{equation}\label{eq:EeuDef}
 E_{\e} (u) := \frac{1}{2 \Card [a,b] \cap \e\Z} \sum_{i \neq j \in \Z \cap \frac{1}{\e} [a,b]} W \left( \frac{u(\e j) - u(\e i)}{\e} \right) ,
\end{equation}
where $W: \R \setminus \{0\} \to \R$ is a two-body interaction potential with suitable decay at infinity (for example, a Lennard-Jones potential).
Next, we do a Taylor expansion of $E_{\e} (u)$ with respect to $\e$, with the hope of recovering the expression \eqref{eq:I}.
To be precise, we compute the first terms of that expansion:
\begin{equation}\label{eq:EeuIntro}
 E_{\e} (u) = E_0 + E_1 \e + E_2 \e^2 + o(\e^2) ,
\end{equation}
and, in this way, we partially succeed in justifying the model \eqref{eq:I}, in the sense that $E_0$ accounts for the elastic energy, $E_1$ accounts for the sharp interface energy (plus boundary terms), and $E_2$ accounts for the smooth interface energy (plus boundary and jump terms).

The essential difference with the paper of Blanc, Le Bris \& Lions \cite{BLL02} is the regularity assumption on $u$.
While in \cite{BLL02} they assumed that $u$ was smooth enough so that all the Taylor expansions made sense, here we assume that $u$ is continuous and piecewise smooth, so that $u'$ can have jumps.
In this way, we obtain, within $E_1$, a term that accounts for the sharp interfaces.
Sections \ref{se:smooth}--\ref{se:separated} are devoted precisely to the computation of \eqref{eq:EeuIntro} from \eqref{eq:EeuDef}.

In Sections \ref{se:repulsion} and \ref{se:optimal} we change slightly the above atomistic model in order to get, in the continuum limit, an additional term that accounts for the repulsion between sharp interfaces.
The motivation of that analysis was led by our desire to obtain the expression of a repulsion term between interfaces from an atomistic energy.
In the continuum context, it is not clear how to define an energy that represents a repulsion term between interfaces, let alone in our setting where we have two kind of interfaces: smooth and sharp.
In general terms, that repulsion energy would be given by a non-local expression depending on the distance between interfaces, and be a decreasing function of that distance which tends to infinity when the distance tends to zero.
It is not clear, however, how to express that in a formula, since a `smooth interface region' is only defined in vague terms as a region where the absolute value of the second derivative is very high.
Among the three possible kinds of interaction between two interfaces (smooth-smooth, smooth-sharp and sharp-sharp), only the sharp-sharp interface interaction energy is easy to model, namely, as a function depending on the distance between the two interfaces, as describe above.
Here we are using the fact that, in dimension $1$, a sharp interface is represented by a single point; in higher dimensions, in contrast, the sharp-sharp interface interaction energy is not so easy to define.

In Section \ref{se:repulsion} we try to derive this repulsion term between sharp interfaces, and, again, we succeed only partially.
The derivation follows a similar procedure to the one described above, but with an important difference that we describe now.
Let $\e>0$ be the interatomic distance.
Then we assume that the two sharp interfaces are located at $t_0 \in (a,b)$ and $t_0 + m \e \in (a,b)$, where $m\geq 2$ is a natural number.
So we are assuming that the interfaces are at a distance which is a fixed multiple of the interatomic distance.
This assumption was motivated by the experiments of Baele, van Tendeloo \& Amelinckx \cite{BVTA87}, who observed a quasiperiodic microtwinning in the alloy Ni-Mn resulting from a Martensitic transformation, and whose images suggest that two consecutive sharp interfaces are separated by a distance which is $6$ or $10$ or $11$ times the interatomic distance.

The procedure is as follows.
As above, we assume that there is a continuous deformation $u: [a,b] \to \R$ which is smooth in $[a,t_0]$ and $[t_0,b]$, for some $t_0 \in (a,b)$, and such that the atomistic deformation $u_{\e}$ follows $u$ in $[a,t_0] \cup [t_0 + \e m ,b]$ (much like in the analysis of the Section \ref{se:sharp}), but follows a discrete deformation $y : \{ t_0 + \e, \ldots, t_0 + (m-1)\e \} \to \R$ in $(t_0, t_0+ m \e)$.
The discrete deformation $y$ will be given by an optimal profile problem.
To be precise, we consider the energy \eqref{eq:EeuDef} of this $u_{\e}$ and calculate its Taylor expansion \eqref{eq:EeuIntro}.
We interpret the difference between this energy and the energy of the first model as a repulsion term between interfaces.
Naturally, the coefficients $E_0, E_1, E_2$ will depend not only on $u$ but also on $y$.
We will see that the coefficient $E_0$ does not depend on $y$.
The coefficient $E_1$ does depend on $y$, and we choose $y$ to minimise $E_1$.
It turns out that, in many cases, the optimal $y$ is the straight line.
This might explain why, in the region between two sharp interfaces, the atoms are aligned in a straight line.
A further minimisation process shows that the optimal $m$ is $6$, which qualitatively coincides with the experiments of Baele, van Tendeloo \& Amelinckx \cite{BVTA87} explained above.

In this paragraph we compare our approaches to similar ones found in the literature.
As mentioned above, the analysis of Section \ref{se:sharp} follows closely that of Blanc, Le Bris \& Lions \cite{BLL02}, with the important difference that our continuum deformation $u$ is continuous and piecewise smooth (instead of being smooth).
The closest to the analysis of Section \ref{se:repulsion} that we have found in the literature is the paper of Blanc \& Le Bris \cite{BLB05}.
They assume that the sharp interfaces are at a distance $\g$ which is larger than the atomistic scale but smaller than macroscopic, so $0<\e \ll \g \ll 1$; this is different from our approach, as our $\g$ would be $m \e$.
Their atomistic energy equals the analogue of \eqref{eq:EeuDef} plus a term accounting for the energy between the two sharp interfaces.
That energy is given again by an optimal profile problem, but, in their case, the deformation between two sharp interfaces is a continuum one.
Although they approach is very natural, it does not predict a repulsion term between interfaces; quite the opposite: the energy associated with two consecutive sharp interfaces is an increasing function of the distance between them.
We believe that that conclusion is wrong, and this was one of our motivations to present our model of Section \ref{se:repulsion}.

Of course, there is a great number of atomic-to-continuum derivations for elastic and surface energy.
We have pointed out those whose approach is similar to that adopted here.
For radically different approaches, see, for example, Arndt \& Griebel \cite{AG05} (who use an upscaling method) and Braides \& Cicalese \cite{BC07} (who use $\G$-convergence), as well as the references therein.

There are several disadvantages to be found in the atomic-to-continuum method used in this paper.
Some of them are already present in Blanc, Le Bris \& Lions \cite{BLL02}: the macroscopic displacement is assumed to be equal to the microscopic one, the atomic deformation is assumed to follow a given smooth (or piecewise smooth, in our case) deformation, the lattice is assumed to be periodic even near the boundary of the body, the Taylor expansion \eqref{eq:EeuIntro} is not uniform in $u$, there is no guarantee that minimisers of the atomistic energy converge to minimisers of the continuum energy, and, finally, the limit continuum problem is not well-posed (because it is not coercive).
In addition, in our approach we also have the disadvantage that, although we allow for more general deformations (piecewise smooth as opposed to smooth), the discontinuity set of the deformation gradient is prescribed.
There are, nevertheless, some advantages to be found in our approach.
First, it detects the right scaling between smooth and sharp interfaces: the term accounting for sharp interfaces appears in the coefficient $\e$ of the Taylor expansion \eqref{eq:EeuIntro}, while the term accounting for smooth interfaces appears in the coefficient $\e^2$; this corroborates the scaling deduced in Ball \& Mora-Corral \cite{BMC08} by $\G$-convergence methods.
Second, it detects (or suggests) the `general form' of a continuum energy functional accounting for elastic energy, sharp-interface energy and smooth-interface energy, thus partially justifying the model of \cite{BMC08}.
Third, it is able to predict a repulsion term between sharp intrefaces and to describe roughly the atomistic configuration between two consecutive sharp interfaces, in qualitative agreement with the experiments of Baele, van Tendeloo \& Amelinckx \cite{BVTA87}.
We do not consider the restriction to the one-dimensional setting to be a disadvantage, because we believe that many of the calculations in this paper can be carried out to the higher-dimensional case, as done in Blanc, Le Bris \& Lions \cite{BLL02}.

The outline of this paper is as follows.
In Section \ref{se:smooth} we introduce the general notation of the paper, and state one of the main results of Blanc, Le Bris \& Lions \cite{BLL02}, which is the Taylor expansion \eqref{eq:EeuIntro} for smooth deformations, and constitutes the starting point of this paper.
In Section \ref{se:map} we study the map that transforms the atomistic potential into the elastic one.
To be precise, a consequence of \cite{BLL02} (which we recall in Section \ref{se:smooth}) is that the term $E_0$ of \eqref{eq:EeuIntro} has the form $\int_a^b W_1 (u')$, for a function $W_1$ depending on $W$.
We show that blow-up rates at $0$ and decay rates at $\infty$ of $W$ imply the corresponding properties on $W_1$, and we give sufficient conditions for the map $W \mapsto W_1$ to be an isomorphism.
Section \ref{se:sharp} is the core of the paper and contains the calculation of the Taylor expansion \eqref{eq:EeuIntro} for a continuous and piecewise smooth deformation $u$ having exactly one sharp interface.
In Section \ref{se:sign} we compare the conclusion of the result of \cite{BLL02} recalled in Section \ref{se:smooth} with our result of Section \ref{se:sharp}.
In our analysis, there is a new term accounting for the sharp interfaces; we study the sign of that term and give sufficient conditions for it to be positive; this positivity physically means that we need energy to create a sharp interface.
Section \ref{se:separated} shows that if the deformation has several sharp interfaces that are separated at a macroscopic distance, then the analysis does not essentially differ from the case of one interface (showed in Section \ref{se:sharp}), since the term accounting for the interaction between sharp interfaces is of order $o(\e^2)$.
Section \ref{se:repulsion} contains the other principal result of the paper: we present the model for a deformation with two sharp interfaces that are separated at a distance multiple of the atomic one, and calculate the corresponding Taylor expansion.
Section \ref{se:optimal} compares the conclusion of the results of Sections \ref{se:sharp} and \ref{se:repulsion}.
The new term that appears in Section \ref{se:repulsion} is a term that accounts for a repulsion force between the two sharp interfaces; in some particular cases, we solve the optimal problem and, thus, describe the atomistic configuration between two consecutive sharp interfaces.

\section{Smooth deformations: no sharp interfaces}\label{se:smooth}

In this section we introduce the general setting and notation of the paper, and state the result on the Taylor expansion of the atomistic energy in the case of a smooth deformation.

It is known that in the atomic-to-continuum analysis, the Taylor expansion as $\e$ goes to zero depends on the particular sequence of $\e$ going to zero (see, e.g., Blanc, Le Bris \& Lions \cite{BLL02} or Braides \& Cicalese \cite{BC07}).
Usually, there is a natural choice of sequence of $\e$ going to zero.
In this paper, we calculate all the possible limits according to the particular sequence of $\e$ going to zero.
In the next two lemmas, we introduce the language and main properties in order to deal with particular sequences of $\e$ going to zero.

\begin{lemma}\label{le:parameters}
For every $a,b,c \in \R$ and $\e>0$ satisfying $a < b$, let $k_1, k_2, N \in \Z$ be defined by the condition
\begin{equation}\label{eq:k1k2N}
 \frac{1}{\e} a - c \leq k_1 < \frac{1}{\e} a - c + 1 , \qquad \frac{1}{\e} b - c -1 < k_2 \leq \frac{1}{\e} b - c , \qquad N = k_2 - k_1 +1 .
\end{equation}
Then $N = \Card \e\Z \cap [a,b]$ and, as $\e \to 0^+$,
\[
 \frac{1}{N} \frac{1}{\e} = \frac{1}{b-a} + O(\e) , \qquad \e (c + k_1) - a = O(\e) , \qquad b - \e (c + k_2) = O(\e).
\]
Now let $\e \to 0^+$ be a sequence such that there exist
\begin{equation}\label{eq:a1a2b1b2c1c2R}
 a_1, a_2, b_1, b_2, c_1, c_2 \in \R
\end{equation}
satisfying
\begin{equation}\label{eq:a1a2b1b2c1c2}
\begin{split}
 & \e (c + k_1) - a = a_1 \e + a_2 \e^2 + o(\e^2) , \qquad  b - \e (c + k_2) = b_1 \e + b_2 \e^2 + o(\e^2) , \\
 & \frac{1}{N} \frac{1}{\e} = \frac{1}{b-a} + c_1 \e + c_2 \e^2 + o(\e^2) ;
\end{split}
\end{equation}
then
\begin{align*}
 & 0 \leq a_1 \leq 1,  \qquad \mbox{if } \ a_1 = 1 \ \mbox{ then } \ a_2 \leq 0 , \qquad  \mbox{if } \ a_1 = 0 \ \mbox{ then } \ a_2 \geq 0 , \\
 & 0 \leq b_1 \leq 1,  \qquad \mbox{if } \ b_1 = 1 \ \mbox{ then } \ b_2 \leq 0 , \qquad \mbox{if } \ b_1 = 0 \ \mbox{ then } \ b_2 \geq 0 , \\
 & |c_1| \leq \frac{1}{(b-a)^2}, \qquad \mbox{if } \ c_1 = \frac{1}{(b-a)^2} \ \mbox{ then } \ c_2 \leq \frac{1}{(b-a)^3} , \\
 & \hspace*{8.7em} \mbox{if } \ c_1 = \frac{-1}{(b-a)^2} \ \mbox{ then } \ c_2 \geq \frac{1}{(b-a)^3} .
\end{align*}
\end{lemma}
\begin{proof}
We have
\[
 \frac{b-a}{\e} - 1 < N \leq \frac{b-a}{\e} + 1 ,
\]
hence
\[
 \frac{-1}{(b-a+\e)(b-a)} \leq \frac{1}{\e} \left( \frac{1}{N} \frac{1}{\e} - \frac{1}{b-a} \right) <
 \frac{1}{(b-a-\e)(b-a)} .
\]
If $c_1 = \frac{1}{(b-a)^2}$ then
\[
 \frac{1}{\e^2} \left( \frac{1}{N} \frac{1}{\e} - \frac{1}{b-a} - c_1 \e \right) < \frac{1}{(b-a-\e)(b-a)^2},
\]
whereas if $c_1 = \frac{-1}{(b-a)^2}$ then
\[
 \frac{1}{\e^2} \left( \frac{1}{N} \frac{1}{\e} - \frac{1}{b-a} - c_1 \e \right) \geq \frac{1}{(b-a+\e)(b-a)^2} .
\]

Similarly,
\[
 \frac{-a_1}{\e} \leq \frac{\e (c+k_1) - a - a_1 \e}{\e^2} < \frac{1-a_1}{\e} ,
\]
and this implies the corresponding properties for $a_1,a_2$.
The proof for $b_1$ and $b_2$ is analogous.
\end{proof}

For the sake of simplicity, many of the theorems in this paper will be stated and proved when the deformation is defined in $[-1,1]$, and the lattice is $\Z$.
For this choice, we can say more about the parameters that appear in Lemma \ref{le:parameters}.

\begin{lemma}\label{le:[-1,1]}
Let $a=-1$, $b=1$ and $c=0$.
For each $\e>0$ define $k_1, k_2, N \in \Z$ by \eqref{eq:k1k2N}.
Then $k_2 = - k_1$ and $N= 2 k_2 +1$.
Now let $\e \to 0^+$ be a sequence such that there exist \eqref{eq:a1a2b1b2c1c2R} satisfying \eqref{eq:a1a2b1b2c1c2}; then
\begin{equation}\label{eq:a1b1[-1,1]}
\begin{split}
 & b_1 = a_1 \in [0,1], \qquad b_2 = a_2 , \qquad c_1 = \frac{a_1}{2} - \frac{1}{4} , \qquad c_2 = \frac{a_1^2}{2} - \frac{a_1}{2} + \frac{a_2}{2}+ \frac{1}{8} , \\
 & \mbox{if } \ a_1 = 1 \ \mbox{ then } \ a_2 \leq 0 , \qquad  \mbox{if } \ a_1 = 0 \ \mbox{ then } \ a_2 \geq 0 .
\end{split}
\end{equation}
Moreover, for every \eqref{eq:a1a2b1b2c1c2R} such that \eqref{eq:a1b1[-1,1]}, there exists a sequence $\e \to 0^+$ such that \eqref{eq:a1a2b1b2c1c2}.
\end{lemma}
\begin{proof}
The equalities $k_2 = - k_1$ and $N= 2 k_2 +1$ follow at once from the definition.

Now take \eqref{eq:a1a2b1b2c1c2R} and a sequence $\e\to0^+$ such that \eqref{eq:a1a2b1b2c1c2}.
The facts $b_1 = a_1$ and $b_2 = a_2$ follow from the equality $k_2 = - k_1$.
We have, successively,
\[
 \e k_2 = 1 - a_1 \e - a_2 \e^2 + o(\e^2) , \qquad N \e = 2 + (1 - 2a_1) \e - 2 a_2 \e^2 + o(\e^2) ,
\]
and using the general formula
\begin{equation}\label{eq:inverseexpansion}
 (p_0 + p_1 \e + p_2 \e^2 + o(\e^2))^{-1} = \frac{1}{p_0} - \frac{p_1}{p_0^2}\e + \frac{p_1^2 - p_0 p_2}{p_0^3} \e^2 + o(\e^2) , \qquad p_0 \in \R \setminus \{0\}, \quad p_1, p_2 \in \R ,
\end{equation}
we obtain the equalities of $c_1$ and $c_2$ of \eqref{eq:a1b1[-1,1]}.
The rest of relations of \eqref{eq:a1b1[-1,1]} follows from Lemma \ref{le:parameters}.

In order to prove the last part of the theorem, we construct, for every $a_1 \in [0,1]$ and $a_2 \in \R$ such that
\[
 \mbox{if } \ a_1 = 1 \ \mbox{ then } \ a_2 \leq 0 , \qquad  \mbox{if } \ a_1 = 0 \ \mbox{ then } \ a_2 \geq 0 ,
\]
the sequence $\{\e_n\}_{n\in\N}$ of positive numbers tending to zero defined by
\[
 \e_n := \left\{
 \begin{array}{lll}
 \frac{n}{n^2 + a_1 n + a_2} & \mbox{ if } & (a_1,a_2) \neq (1,0) , \\[1ex]
 \frac{n^2}{n^3 + n^2 -1} & \mbox{ if } & (a_1,a_2) = (1,0) 
 \end{array}
 \right. \qquad n \in \N .
\]
The corresponding sequence $\{k_1^n\}_{n\in\N}$ of $k_1$ satisfies $k_1^n = -n$ for big $n \in \N$, and $\e_n k_1^n + 1 = a_1 \e_n + a_2 \e_n + o(\e_n^2)$.
\end{proof}

With all these preliminaries, we are in a position to present the model of the atomistic energy of a deformation of an elastic crystal.
The rest of the paper will be, of course, devoted to the analysis of that model.

The one-dimensional elastic body is represented, in the reference configuration, by the closed interval $[a,b]$, for some real numbers $a<b$.
Although it is customary to represent the reference configuration through an open set, in our case, we believe that the calculations are slightly simpler; in any case, it makes very little difference.
The continuum deformation of the body is represented by a continuous increasing map $u: [a,b] \to \R$.
The continuity models that no fracture is allowed, and being increasing models the orientation-preserving character of the deformation and the non-interpenetration of matter.
We assume that the body possesses a crystalline structure; in particular, we choose the lattice $\ell := c + \Z$ for some $c \in \R$, we take $\e>0$ as the interatomic distance (which in the end will go to zero) and assume that the atoms of the body are located at the points of $\e \ell \cap [a,b]$.
Thus, the number of atoms of the body is $\Card \e \ell \cap [a,b]$, which, in the notation of Lemma \ref{le:parameters}, coincides with $N$.
We assume that the atomistic deformation $u_{\e} : \e \ell \cap [a,b] \to \R$ is the restriction to $\e \ell \cap [a,b]$ of the continuum deformation $u$.
The atomistic energy of the discrete deformation $u_{\e}$ is given by a two-body interaction potential.
This assumption is known to be very simplistic, but we believe that a good understanding of this model is needed prior to the analysis of more general and realistic ones.
So let $W: \R \setminus \{0\} \to \R$ be the two-body interaction potential, which is a continuous function with some decay properties at infinity; the precise assumption will be stated in Theorem \ref{th:2ndorder} below.
We assume that the atomistic displacement equals the macroscopic one, and, hence, we define the atomistic energy of the deformation $u_{\e}$ as
\begin{equation}\label{eq:Eeue}
   E_{\e} (u_{\e}) := \frac{1}{2 \Card \e \ell \cap [a,b]} \sum_{i\neq j \in \ell \cap \frac{1}{\e} [a,b]} W \left( \frac{u_{\e}(\e j) - u_{\e}(\e i)}{\e} \right) .
\end{equation}
Note that, in terms of the notation \eqref{eq:k1k2N}, the energy can be equivalently written as
\[
 E_{\e} (u_{\e}) = \frac{1}{2 N} \sum_{\substack{i,j= 0 \\ i\neq j}}^{k_2 - k_1} W \left( \frac{u(\e(c+k_1+j)) - u(\e(c+k_1+i))}{\e} \right) .
\]

The above paragraph has described, essentially, the particularisation to dimension $1$ of the model presented by Blanc, Le Bris \& Lions \cite{BLL02}.
The only modification is that, in our case, the body is represented by a closed interval (not an open one) and we are working with an arbitrary sequence of $\e \to 0^+$.
Because of those (minor) modifications, the following result is not a particular case of Blanc, Le Bris \& Lions \cite[Th.\ 3]{BLL02}, but, since the proof follows exactly the same lines (and in fact, it is simpler because of the 1D assumption), we omit it.

\begin{theorem}\label{th:2ndorder}
Let $W : \R \setminus \{0\} \to \R$ be a $\mc{C}^{\infty}$ function such that $W(x) = W(-x)$ for all $x \in \R \setminus \{0\}$, and satisfy that there exist $C,R >0$ and $\a>3$ such that 
\begin{equation}\label{eq:assumptionsW}
 |W^{i)} (x)| \leq C |x|^{-\a-i} , \quad \mbox{for all} \quad x \in \R \setminus (-R,R) \quad \mbox{and} \quad i \in \{0,1,2,\ldots\}.
\end{equation}
Let $a,b,c \in \R$ and $\e>0$ with $a<b$.
Define $\ell := c + \Z$.
Let $u : [a,b] \to \R$ be a $\mc{C}^{\infty}$ diffeomorphism.
Let $u_{\e}$ be the restriction of $u$ to $\e \ell \cap [a,b]$
Take \eqref{eq:a1a2b1b2c1c2R} and a sequence $\e \to 0^+$ such that \eqref{eq:a1a2b1b2c1c2}.
Define \eqref{eq:Eeue}.
Then
\begin{equation}\label{eq:Eeu}
 E_{\e} (u_{\e}) = E^0 + \e E^1 + \e^2 E^2 + o(\e^2) ,
\end{equation}
where
\begin{equation}\label{eq:E0}
  E^0 := \frac{1}{b-a} \int_a^b \sum_{j=1}^{\infty} W(u'(x)j) \dd x ,
\end{equation}
\begin{equation}\label{eq:E1}
 E^1 := c_1 (b-a) E^0 - \frac{1}{2(b-a)} \sum_{j=1}^{\infty} \left[ (j + 2 a_1 -1) W(u'(a)j) + (j + 2 b_1 - 1) W(u'(b)j) \right] ,
\end{equation}
\begin{equation}\label{eq:E2}
\begin{split}
 E^2 := & - \frac{1}{24(b-a)} \int_a^b  \sum_{j=1}^{\infty} W''(u'(x)j) u''(x)^2 j^4 \dd x
 + c_2 (b-a) E_0
 \\ & + \sum_{j=1}^{\infty} \left( \frac{c_1}{2} - a_1 c_1 - \frac{a_2}{b-a} - \frac{c_1}{2}j \right) W(u'(a)j) \\
 & + \sum_{j=1}^{\infty} \left( \frac{c_1}{2} - b_1 c_1 - \frac{b_2}{b-a} - \frac{c_1}{2}j \right) W(u'(b)j)
 \\ & + \frac{1}{b-a} \sum_{j=1}^{\infty} 
        \left[ \left( - \frac{1}{12} + \frac{a_1}{2} - \frac{a_1^2}{2} \right) j + \left( \frac{1}{4} - \frac{a_1}{2} \right) j^2 - \frac{1}{6}j^3 \right] W'(u'(a)j) u''(a)
 \\ & + \frac{1}{b-a} \sum_{j=1}^{\infty} \left[ \left( \frac{1}{12} - \frac{b_1}{2} + \frac{b_1^2}{2} \right) j + \left( - \frac{1}{4} + \frac{b_1}{2} \right) j^2 + \frac{1}{6}j^3 \right] W'(u'(b)j) u''(b) .
\end{split}
\end{equation}
\end{theorem}

Theorem \ref{th:2ndorder} requires some regularity of $u$ and $W$, and decay conditions on $W$ and its derivatives.
As explained in the proof of Blanc, Le Bris \& Lions \cite[Th.\ 3]{BLL02}, it is possible to prove \eqref{eq:Eeu} with a different set of hypotheses on $u$ and $W$; roughly speaking, less regularity of $u$ requires stronger decay conditions on $W$.
In fact, a proof of a version of Theorem \ref{th:2ndorder} with weaker assumptions on $u$ and $W$ was made by Blanc \& Le Bris \cite[Th.\ 2.1]{BLB05}.
However, in this paper we are not interested in relaxing the regularity assumptions on $u$.
Note also that, in Theorem \ref{th:2ndorder}, we have written explicitly the boundary term in \eqref{eq:E2}, which was not done in the higher-dimensional case of Blanc, Le Bris \& Lions \cite{BLL02}.

\section{The map that transforms the atomistic into the elastic potential}\label{se:map}

In Nonlinear Elasticity Theory (see, e.g., Ball \cite{Ball77}), it is assumed the existence of a function $W_1 : \R \to \R \cup \{\infty\}$, called the (elastic) stored-energy function of the material, such that the elastic energy of a deformation $u : (a,b) \to \R$ of the body represented by the interval $(a,b)$ is $\int_a^b W_1(u'(x)) \mathrm{d} x$ (of course, we have restricted the general theory to the one-dimensional case).
This and Theorem \ref{th:2ndorder} suggest the introduction of the operator that maps any function $W: (0,\infty) \to \R$ into the function $W_1$ defined by $W_1 (t) := \sum_{j=1}^{\infty} W(jt)$, for each $t>0$ for which the series converges.
In this section we give a sufficient condition for that operator transforming the atomistic potential $W$ into the (continuum) elastic one $W_1$ to be an isomorphism.

Let $p,q \in \R$.
Define $\mc{A}_{p,q}$ as the set of $f \in \mc{C}(0,\infty)$ such that $\limsup_{t \to 0+} t^p |f(t)| < \infty$ and $\limsup_{t \to \infty} t^q |f(t)| < \infty$.
Clearly, $\mc{A}_{p,q}$ is a vector space.
For each $f \in \mc{A}_{p,q}$, define
\begin{equation*}
 \| f \|_{p,q} := \max \left\{ \sup_{t \in (0,1)} t^p |f(t)| , \sup_{t \in [1,\infty)} t^q |f(t)| \right\} .
\end{equation*}
It is immediate to see that $\| \cdot \|_{p,q}$ is a norm in $\mc{A}_{p,q}$ that equips it with the structure of Banach space.

In this section (and also in Section \ref{se:sign}), we let $\zeta : (1,\infty) \to \R$ denote the restriction to $(1,\infty)$ of Riemann's zeta function, i.e., $\zeta (s) := \sum_{j=1}^{\infty} j^{-s}$ for each $s>1$.
We denote the norm of a linear operator between two Banach spaces simply as $\|\cdot\|$; the identity operator is denoted by $I$.
Motivated by the introduction of this section, for each $k \in \N$ we consider the operator $T_k$ that maps any function $f:(0,\infty) \to \R \cup \{\infty\}$ into the function $T_k f$ defined by $T_k f (t) := \sum_{j=1}^{\infty} j^k f(j t)$ for each $t >0$ for which that series converges (to a number or to $\infty$).

\begin{lemma}\label{le:Tiso}
Let $k \in \N$ and $p,q> k +1$.
Then $T_k : \mc{A}_{p,q} \to \mc{A}_{p,q}$ is a linear bounded operator.
If $p \geq q$ then $\| T_k - I \| \leq \zeta(q-k) -1$.
\end{lemma}
\begin{proof}
Call $S_k := T_k - I$.
For each $t \geq 1$ and $f \in \mc{A}_{p,q}$ we have $t^q \left| S_k f (t) \right| \leq \left( \zeta(q-k) - 1 \right) \| f \|_{p,q}$.
For each $0 < t < 1$, define $j_t$ as the only integer satisfying $j_t t \geq 1$ and $(j_t - 1) t < 1$.
Call $A_t:= \sum_{j=2}^{j_t -1} j^{k-p} + t^{p-q} \sum_{j=j_t}^{\infty} j^{k-q}$ and $A:= \sup_{t \in (0,1)} A_t$.
Then $t^p \left| S_k f (t) \right| \leq A_t \| f \|_{p,q}$.
If we prove that $A< \infty$, then we will have shown that $S_k$ is a linear bounded operator with $\| S_k \| \leq \max \{ \zeta(q-k) - 1 , A \}$.

Now we prove that $A < \infty$.
If $p\geq q$ then $A \leq \zeta(q-k) - 1$, whereas if $p< q$ then
\[
 \sup_{t \in [\frac{1}{2}, 1)} A_t \leq \sup_{t \in [\frac{1}{2} , 1)} \left[ \sum_{j=2}^{j_t -1} j^{k-p} + 2^{q-p} \sum_{j=j_t}^{\infty} j^{k-p} \right] \leq 2^{q-p} (\zeta(p-k) - 1)
\]
and
\[
 \sup_{t \in (0, \frac{1}{2} )} A_t \leq \zeta(p-k) - 1 + \sup_{t \in (0,\frac{1}{2} )} t^{p-q} \int_{j_t - 1}^{\infty} s^{k-q} \dd s \leq \zeta(p-k) - 1 + \frac{2^{q-p}}{q-k-1} .
\]
Therefore, $A< \infty$.
\end{proof}

Let $k \in \N$.
For each $i \in \{ 0, \ldots, k \}$ let $p_i, q_i \in \R$, and define $p:=(p_0, \ldots, p_k)$ and $q:=(q_0, \ldots, q_k)$.
Let $\mc{A}_{p,q;k}$ be the Banach space of functions $f \in \mc{C}^k (0, \infty)$ such that $f^{i)} \in \mc{A}_{p_i, q_i}$ for each $i \in \{ 0, \ldots, k \}$, equipped with the norm $\| f \|_{p,q;k} := \sum_{i=0}^k \| f^{i)} \|_{p_i,q_i}$.

\begin{proposition}\label{prop:iso}
Let $k \in \N$.
For each $i \in \{ 0, \ldots, k \}$, consider $p_i, q_i > i+1$.
Define $p:=(p_0, \ldots, p_k)$ and $q:=(q_0, \ldots, q_k)$.
Then $T_0 : \mc{A}_{p,q;k} \to \mc{A}_{p,q;k}$ is a bounded linear operator.
If $p_i \geq q_i > \zeta^{-1} (2) + i$ for all $i \in \{ 0, \ldots, k \}$, then $T_0 : \mc{A}_{p,q;k} \to \mc{A}_{p,q;k}$ is an isomorphism.
\end{proposition}
\begin{proof}
Let $f \in \mc{A}_{p,q;k}$.
By Lemma \ref{le:Tiso},
\[
  \| T_0 f \|_{p,q;k}  = \sum_{i=0}^k \| (T_0 f)^{i)} \|_{p_i,q_i} = \sum_{i=0}^k \| T_i f^{i)} \|_{p_i,q_i} 
 \leq \sum_{i=0}^k \| T_i \| \| f^{i)} \|_{p_i,q_i} \leq \max_{0 \leq i \leq k} \|T_i\| \| f \|_{p,q;k} .
\]
If $p_i \geq q_i > \zeta^{-1} (2) + i$ for all $i \in \{ 0, \ldots, k \}$, then, again by Lemma \ref{le:Tiso}, $\| T_0 - I \| \leq \max_{0 \leq i \leq k} \|T_i - I\| \leq \max_{0 \leq i \leq k} \zeta(q_i-i)-1 < 1$, and, hence, $T_0$ is an isomorphism.
\end{proof}

Note that $\zeta^{-1} (2) \simeq 1.72865$.
As an example, motivated by the Lennard-Jones potential (see \eqref{eq:LJ} below), for each $k\in\N$ define $LJ_k := \mc{A}_{(12, \ldots, 12+k),(6, \ldots,6+k);k}$.
Then, the Lennard-Jones potential \eqref{eq:LJ} belongs to $LJ_k$, and by Proposition \ref{prop:iso}, $T_0 : LJ_k \to LJ_k$ is an isomorphism.

Finally, we recall that Ventevogel \cite{Ventevogel78} constructed an example of a continuous function $\phi : (0,\infty) \to \R \cup \{\infty\}$ with exactly one relative minimum, and such that $T_0 \phi$ has several relative minima.
In fact, his example can be easily adapted to construct, for each $p,q>1$, a smooth function $W \in \mc{A}_{p,q}$ such that $W$ has exactly one relative minimum, and $T_0 W$ has several relative minima.

\section{Piecewise smooth deformations: sharp interfaces}\label{se:sharp}

This section is devoted to the proof of the following result, which is the analogue of Theorem \ref{th:2ndorder} for deformations that are continuous and piecewise smooth.

\begin{theorem}\label{th:2ndorderjump}
Let $W : \R \setminus \{0\} \to \R$ be a $\mc{C}^{\infty}$ function such that $W(x) = W(-x)$ for all $x \in \R \setminus \{0\}$, and satisfy that there exist $C,R >0$ and $\a>3$ such that \eqref{eq:assumptionsW}.
Fix $a=-1$ and $b=1$.
Define $\ell := \Z$.
Let $u : [-1,1] \to \R$ be continuous, increasing and satisfy that $u|_{[-1,0]}$ and $u|_{[0,1]}$ are $\mc{C}^{\infty}$ diffeomorphisms.
For each $\e>0$, let $u_{\e}$ be the restriction of $u$ to $\e \ell \cap [a,b]$
Take \eqref{eq:a1a2b1b2c1c2R} and a sequence $\e \to 0^+$ such that \eqref{eq:a1a2b1b2c1c2}.
Define \eqref{eq:Eeue}.
Then \eqref{eq:Eeu}, where
\begin{equation*}
  E^0 := \frac{1}{2} \int_{-1}^1 \sum_{j=1}^{\infty} W(u'(x)j) \dd x ,
\end{equation*}
\begin{align*}
 E^1 := & \left( a_1 - \frac{1}{2} \right) E^0
 - \frac{1}{4} \sum_{j=1}^{\infty} (j + 2 a_1 -1) \left[ W(u'(-1)j) + W(u'(1)j) \right] \\
 & - \frac{1}{4} \sum_{j=2}^{\infty} (j - 1) \left[ W(u'(0^-)j) + W(u'(0^+)j) \right]
 + \frac{1}{2} \sum_{i,j= 1}^{\infty} W \left( u'(0^+) j + u'(0^-) i \right) ,
\end{align*}
\begin{align*}
 & E^2 := - \frac{1}{48} \int_{-1}^1 \sum_{j=1}^{\infty} W''(u'(x)j) u''(x)^2 j^4 \dd x 
 + \left( \frac{1}{4} - a_1 + a_2 + a_1^2 \right) E^0
 \\ & + \frac{1}{2} \sum_{j=1}^{\infty} \left[ - \frac{1}{4} + a_1 - a_2 - a_1^2 + \left( \frac{1}{4} - \frac{a_1}{2} \right) j \right] \left[ W(u'(-1)j) + W(u'(1)j) \right]
 \\ & + \frac{1}{4} \sum_{j=1}^{\infty} \!
        \left[ \left( \frac{1}{6} -a_1 + a_1^2 \right) \! j + \! \left( a_1-\frac{1}{2} \right)\!  j^2 + \frac{1}{3}j^3 \right] \! \left[ W'(u'(1)j) u''(1) - W'(u'(-1)j) u''(-1) \right]
 \\ & + \left( \frac{1}{8} - \frac{a_1}{4} \right) \sum_{j=2}^{\infty} (j-1) \left[ W(u'(0^-)j) + W(u'(0^+)j) \right]
 \\ & + \frac{1}{4} \sum_{j=2}^{\infty} 
        \left( \frac{1}{6}j - \frac{1}{2} j^2 + \frac{1}{3}j^3 \right) \left[ W'(u'(0^-)j) u''(0^-) - W'(u'(0^+)j) u''(0^+) \right]
 \\ & + \left( \frac{a_1}{2} - \frac{1}{4} \right) \sum_{i,j= 1}^{\infty} W \left( u'(0^+) j + u'(0^-) i \right) 
 \\ &  + \frac{1}{4} \sum_{i,j= 1}^{\infty} \left[ u''(0^+) j^2 - u''(0^-) i^2 \right] W' \left( u'(0^+) j + u'(0^-) i \right) .
\end{align*}
\end{theorem}
\begin{proof}
In \eqref{eq:k1k2N}, given $a,b,c \in \R$ and $\e>0$ with $a<b$, we defined $k_1$, $k_2$ and $N$.
In our case, we have $a=-1$, $b=1$ and $c=0$, so define $k_1$, $k_2$, $N$ accordingly.
Define $a^-=-1$, $b^-=0$ and $c^-=0$, and construct $k_1^-$, $k_2^-$, $N^-$ accordingly; finally, define $a^+=0$, $b^+=1$ and $c^+=0$, and construct $k_1^+$, $k_2^+$, $N^+$ accordingly.
It is easy to see that $k^-_1 = k_1$, $k^-_2=0$, $k^+_1=0$ and $k^+_2=k_2$, hence $N^-= N-k_2$ and $N^+= N+k_1$.
By Lemma \ref{le:[-1,1]}, $k_2 = - k_1$, $N= 2 k_2 +1$, $N^- = N^+ = \frac{N+1}{2}$ and \eqref{eq:a1b1[-1,1]}.
Define
\begin{equation}\label{eq:a1-a1+}
\begin{split}
 & a^-_1 := a_1, \quad a^-_2 := a_2, \quad b^-_1 := 0, \quad b^-_2 := 0, \quad a^+_1 := 0, \quad a^+_2 := 0, \quad b^+_1 := b_1 ,\\
 & b^+_2 := b_2, \quad c^-_1 := 2 c_1 - \frac{1}{2}, \quad c^+_1 := c^-_1 , \quad  c^-_2 := -2 c_1 + 2 c_2 + \frac{1}{4}, \quad c^+_2 := c^-_2 .
\end{split}
\end{equation}
Then, straightforward calculations using Lemma \ref{le:[-1,1]} and Formula \eqref{eq:inverseexpansion} show that
\begin{align*}
 & \e (c^{\pm} + k_1^{\pm}) - a^{\pm} = a_1^{\pm} \e + a_2^{\pm} \e^2 + o(\e^2) , \qquad  b^{\pm} - \e (c^{\pm} + k_2^{\pm}) = b_1^{\pm} \e + b_2^{\pm} \e^2 + o(\e^2) , \\
 & \frac{1}{N^{\pm}} \frac{1}{\e} = \frac{1}{b^{\pm}-a^{\pm}} + c_1^{\pm} \e + c_2^{\pm} \e^2 + o(\e^2) .
\end{align*}
Define
\begin{equation*}
 E^{\pm}_{\e} (u_{\e}) := \frac{1}{2 N^{\pm}} \sum_{\substack{i,j= 0 \\ i\neq j}}^{k^{\pm}_2 - k^{\pm}_1} W \left( \frac{u(\e(k^{\pm}_1+j)) - u(\e(k^{\pm}_1+i))}{\e} \right) .
\end{equation*}
By Theorem \ref{th:2ndorder} and \eqref{eq:a1-a1+}, $E^{\pm}_{\e} (u) = (E^0)^{\pm} + \e (E^1)^{\pm} + \e^2 (E^2)^{\pm} + o(\e^2)$, where
\[
 (E^0)^- := \int_{-1}^0 \sum_{j=1}^{\infty} W(u'(x)j) \dd x , \qquad (E^0)^+ := \int_0^1 \sum_{j=1}^{\infty} W(u'(x)j) \dd x ,
\]
\begin{equation*}
 (E^1)^- := (a_1 - 1) (E^0)^-
 - \frac{1}{2} \sum_{j=1}^{\infty} (j + 2 a_1 -1) W(u'(-1)j) 
 - \frac{1}{2} \sum_{j=1}^{\infty} (j - 1) W(u'(0^-)j) ,
\end{equation*}
\[
  (E^1)^+ := (a_1 - 1) (E^0)^+ - \frac{1}{2} \sum_{j=1}^{\infty} (j-1) W(u'(0^+)j) - \frac{1}{2} \sum_{j=1}^{\infty} (j + 2 a_1 - 1) W(u'(1)j) ,
\]
\begin{align*}
 (E^2)^- := & - \frac{1}{24} \int_{-1}^0  \sum_{j=1}^{\infty} W''(u'(x)j) u''(x)^2 j^4 \dd x
 + ( 1 - 2a_1 + a_2 + a_1^2 ) (E^0)^-
 \\ & + \sum_{j=1}^{\infty} \left[ - \frac{1}{2} + \frac{3 a_1}{2} - a_2 - a_1^2 + \left( \frac{1}{2} - \frac{a_1}{2} \right) j \right] W(u'(-1)j)
 \\ & + \sum_{j=1}^{\infty} \left[ - \frac{1}{2} + \frac{a_1}{2} + \left( \frac{1}{2} - \frac{a_1}{2} \right) j \right] W(u'(0^-)j)
 \\ & + \sum_{j=1}^{\infty} 
        \left[ \left( - \frac{1}{12} + \frac{a_1}{2} - \frac{a_1^2}{2} \right) j + \left( \frac{1}{4} - \frac{a_1}{2} \right) j^2 - \frac{1}{6}j^3 \right] W'(u'(-1)j) u''(-1)
 \\ & + \sum_{j=1}^{\infty} \left( \frac{1}{12} j - \frac{1}{4} j^2 + \frac{1}{6}j^3 \right) W'(u'(0^-)j) u''(0^-)
 ,
\end{align*}
\begin{align*}
 (E^2)^+ := & - \frac{1}{24} \int_0^1 \sum_{j=1}^{\infty} W''(u'(x)j) u''(x)^2 j^4 \dd x
 + \left( 1 - 2a_1 + a_2 + a_1^2 \right) (E^0)^+
 \\ & + \sum_{j=1}^{\infty} \left[ - \frac{1}{2} + \frac{a_1}{2} + \left( \frac{1}{2} - \frac{a_1}{2} \right) j  \right] W(u'(0^+)j)
 \\ & + \sum_{j=1}^{\infty} \left[ - \frac{1}{2} + \frac{3a_1}{2} - a_2 - a_1^2 + \left( \frac{1}{2} - \frac{a_1}{2} \right) j  \right] W(u'(1)j)
 \\ & + \sum_{j=1}^{\infty} 
        \left( - \frac{1}{12} j + \frac{1}{4} j^2 - \frac{1}{6}j^3 \right) W'(u'(0^+)j) u''(0^+)
 \\ & + \sum_{j=1}^{\infty} \left[ \left( \frac{1}{12} - \frac{a_1}{2} + \frac{a_1^2}{2} \right) j + \left( - \frac{1}{4} + \frac{a_1}{2} \right) j^2 + \frac{1}{6}j^3 \right] W'(u'(1)j) u''(1) .
\end{align*}

We express
\begin{equation}\label{eq:ps1}
 E_{\e} (u_{\e}) = \frac{1}{2 N} \left[ \sum_{\substack{i,j= 0 \\ i\neq j}}^{- k_1} + \sum_{\substack{i,j= -k_1+1 \\ i\neq j}}^{k_2 - k_1} + 2 \sum_{i= 0}^{- k_1} \sum_{j= -k_1+1}^{k_2- k_1} \right] W \left( \frac{u(\e(k_1+j)) - u(\e(k_1+i))}{\e} \right) .
\end{equation}
Clearly,
\[
 \frac{1}{2 N} \sum_{\substack{i,j= 0 \\ i\neq j}}^{- k_1} W \left( \frac{u(\e(k_1+j)) - u(\e(k_1+i))}{\e} \right) = \frac{N+1}{2N} E^-_\e (u) ,
\]
and, by \eqref{eq:a1a2b1b2c1c2}, \eqref{eq:a1b1[-1,1]} and \eqref{eq:inverseexpansion},
\[
 \frac{N+1}{2N} = \frac{1}{2} + \frac{1}{4} \e + \left( \frac{a_1}{4} - \frac{1}{8} \right) \e^2 + o(\e^2) .
\]
Therefore,
\begin{equation}\label{eq:ps2}
\begin{split}
 \frac{1}{2 N}  \sum_{\substack{i,j= 0 \\ i\neq j}}^{- k_1} W  & \left( \frac{u(\e(k_1+j)) - u(\e(k_1+i))}{\e} \right) 
 = \frac{1}{2} (E^0)^-
 + \e \left[ \frac{1}{4} (E^0)^- + \frac{1}{2} (E^1)^- \right]
 \\ & + \e^2 \left[ \left( \frac{a_1}{4} - \frac{1}{8} \right) (E^0)^- + \frac{1}{4} (E^1)^- + \frac{1}{2} (E^2)^- \right] + o(\e^2) .
\end{split}
\end{equation}

Now we observe that
\begin{multline*}
 \frac{1}{2 N} \sum_{\substack{i,j= -k_1+1 \\ i\neq j}}^{k_2 - k_1} W \left( \frac{u(\e(k_1+j)) - u(\e(k_1+i))}{\e} \right) \\ 
 = \frac{N+1}{2N} E^+_{\e} (u_{\e}) - \frac{1}{N} \sum_{j= 1}^{k_2} W \left( \frac{u(\e j) - u(0)}{\e} \right)
\end{multline*}
with
\begin{align*}
 \frac{N+1}{2N} E^+_{\e} (u_{\e}) = & \frac{1}{2} (E^0)^+ 
 + \e \left[ \frac{1}{4} (E^0)^+ + \frac{1}{2} (E^1)^+ \right] \\
  & + \e^2 \left[ \left( \frac{a_1}{4} - \frac{1}{8} \right) (E^0)^+ + \frac{1}{4} (E^1)^+ + \frac{1}{2} (E^2)^+ \right] + o(\e^2) .
\end{align*}

By Taylor expansion, for each $j \in \{1, \ldots, k_2\}$,
\begin{align*}
 W & \left( \frac{u(\e j) - u(0)}{\e} \right) \\ 
 & = W(u'(0^+)j) + W'(u'(0^+)j) \left( \frac{u(\e j) - u(0)}{\e} - u'(0^+)j \right) + O(\e^2) j^{4-a_2} 
 \\ & = W(u'(0^+)j) +  \e \frac{1}{2} W'(u'(0^+)j) u''(0^+) j^2 + O(\e^2) \left( j^{3-a_1} + j^{4-a_2} \right) .
\end{align*}
Therefore,
\[
 \frac{1}{N} \sum_{j= 1}^{k_2} W \left( \frac{u(\e j) - u(0)}{\e} \right)
 = \frac{1}{N} \sum_{j= 1}^{k_2} W(u'(0^+)j) + \frac{1}{N} \sum_{j= 1}^{k_2} \e \frac{1}{2} W'(u'(0^+)j) u''(0^+) j^2 + o(\e^2) .
\]
It is easy to see that
\[
 \sum_{j= 1}^{k_2} W(u'(0^+)j) = \sum_{j= 1}^{\infty} W(u'(0^+)j) + o(\e), \quad
 \sum_{j= 1}^{k_2} W'(u'(0^+)j) j^2 = \sum_{j= 1}^{\infty} W'(u'(0^+)j) j^2 + o(1) .
\]
Thus,
\begin{align*}
 \frac{1}{N} \sum_{j= 1}^{k_2} W & \left( \frac{u(\e j) - u(0)}{\e} \right) 
 = \e \frac{1}{2} \sum_{j= 1}^{\infty} W(u'(0^+)j)
 \\ & + \e^2 \left[ \left( \frac{a_1}{2} - \frac{1}{4} \right) \sum_{j= 1}^{\infty} W(u'(0^+)j) + \frac{1}{4} \sum_{j= 1}^{\infty} W'(u'(0^+)j) u''(0^+) j^2 \right] + o(\e^2) .
\end{align*}
In total,
\begin{equation}\label{eq:ps3}
\begin{split}
 & \frac{1}{2 N} \sum_{\substack{i,j= -k_1+1 \\ i\neq j}}^{k_2 - k_1} W \left( \frac{u(\e(k_1+j)) - u(\e(k_1+i))}{\e} \right) 
 \\ & = \frac{1}{2} (E^0)^+ 
  + \e \left[ \frac{1}{4} (E^0)^+ + \frac{1}{2} (E^1)^+ - \frac{1}{2} \sum_{j= 1}^{\infty} W(u'(0^+)j)\right]
 \\ & \quad + \e^2 \Bigg[ \left( \frac{a_1}{4} - \frac{1}{8} \right) (E^0)^+ + \frac{1}{4} (E^1)^+ + \frac{1}{2} (E^2)^+ +\left( - \frac{a_1}{2} + \frac{1}{4} \right) \sum_{j= 1}^{\infty} W(u'(0^+)j) 
 \\ & \qquad \qquad  - \frac{1}{4} \sum_{j= 1}^{\infty} W'(u'(0^+)j) u''(0^+) j^2 \Bigg] + o(\e^2) .
\end{split}
\end{equation}

Take $0 \leq i\leq -k_1$ and $-k_1 +1 \leq j \leq k_2 - k_1$. Then
\begin{multline*}
 \frac{u(\e(k_1+j)) - u(\e(k_1+i))}{\e} = u'(0^+) (k_1+j) + u'(0^-) (-k_1-i) 
 \\ + \frac{\e}{2} \left[ u''(0^+) (k_1+j)^2 - u''(0^-) (k_1+i)^2 \right] + O(\e^2) \left[ (k_1+j)^3 + (-k_1-i)^3 \right] .
\end{multline*}
Therefore,
\begin{align*}
 W & \left( \frac{u(\e(k_1+j)) - u(\e(k_1+i))}{\e} \right) = W \left( u'(0^+) (k_1+j) + u'(0^-) (-k_1-i) \right)
 \\ & + W' \left( u'(0^+) (k_1+j) + u'(0^-) (-k_1-i) \right) \frac{\e}{2} \left[ u''(0^+) (k_1+j)^2 - u''(0^-) (k_1+i)^2 \right]
 \\ & + O(\e^2) \left( (j-i)^{-a_1} \left[ (k_1+j)^3 + (-k_1-i)^3 \right] + (j-i)^{-a_2} \left[ (k_1+j)^2 + (-k_1-i)^2 \right]^2 \right) .
\end{align*}
Now
\begin{align*}
  \frac{\e^2}{N} & \sum_{i= 0}^{- k_1} \sum_{j= -k_1+1}^{k_2- k_1} \! \! \left( \! (j-i)^{-a_1} \left[ (k_1+j)^3 + (-k_1-i)^3 \right] + (j-i)^{-a_2} \left[ (k_1+j)^2 + (-k_1-i)^2 \right]^2 \right)
 \\ & =  \frac{1}{N} \e^2 \sum_{i= 0}^{- k_1} \sum_{j= 1}^{k_2} \left[ (j+i)^{-a_1} (j^3 + i^3) + (j+i)^{-a_2} (j^2 + i^2)^2 \right] 
 \\ & \leq C_1 \e^3 \sum_{i= 0}^{- k_1} \sum_{j= 1}^{k_2} \left[ (j+i)^{3-a_1} + (j+i)^{4-a_2} \right]
 \leq C_2 \e^3 \left[ 1 + \sum_{i= 1}^{- k_1} \left( i^{4-a_1} + i^{5-a_2} \right) \right] = o(\e^2) ,
\end{align*}
for some constants $C_1, C_2 >0$ depending on $W,u$, but not on $i,j,\e$.
It is easy to see that
\begin{equation*}
 \sum_{i= 0}^{- k_1} \sum_{j= -k_1+1}^{k_2- k_1} \! W \left( u'(0^+) (k_1+j) + u'(0^-) (-k_1-i) \right) 
  = \sum_{i=0}^{\infty} \sum_{j= 1}^{\infty}W \left( u'(0^+) j + u'(0^-) i \right) + o(\e) .
\end{equation*}
Therefore,
\begin{align*}
 \frac{1}{N} & \sum_{i= 0}^{- k_1} \sum_{j= -k_1+1}^{k_2- k_1} W \left( u'(0^+) (k_1+j) + u'(0^-) (-k_1-i) \right) 
 \\ & = \left[ \frac{1}{2} \e + \left( \frac{a_1}{2} - \frac{1}{4} \right) \e^2 \right] \sum_{i= 0}^{\infty} \sum_{j= 1}^{\infty} W \left( u'(0^+) j + u'(0^-) i \right) + o(\e^2) .
\end{align*}
Similarly, it is also easy to see that
\begin{align*}
 \frac{\e}{2N} \sum_{i= 0}^{- k_1} & \sum_{j= -k_1+1}^{k_2- k_1} \! \! W' \left( u'(0^+) (k_1+j) + u'(0^-) (-k_1-i) \right) \! \left[ u''(0^+) (k_1+j)^2 - u''(0^-) (k_1+i)^2 \right]
 \\ & = \e^2 \frac{1}{4} \sum_{i= 0}^{\infty} \sum_{j= 1}^{\infty}W' \left( u'(0^+) j + u'(0^-) i \right) \left[ u''(0^+) j^2 - u''(0^-) i^2 \right] + o(\e^2) .
\end{align*}
In total,
\begin{equation}\label{eq:ps4}
\begin{split}  \frac{1}{N} & \sum_{i= 0}^{- k_1} \sum_{j= -k_1+1}^{k_2- k_1} W \left( \frac{u(\e(k_1+j)) - u(\e(k_1+i))}{\e} \right) 
 = \e \frac{1}{2} \sum_{i= 0}^{\infty} \sum_{j= 1}^{\infty}W \left( u'(0^+) j + u'(0^-) i \right)
 \\ &
  + \e^2  \Bigg[ \left( \frac{a_1}{2} - \frac{1}{4} \right) \sum_{i= 0}^{\infty} \sum_{j= 1}^{\infty}W \left( u'(0^+) j + u'(0^-) i \right) 
   \\ & \qquad + \frac{1}{4} \sum_{i= 0}^{\infty} \sum_{j= 1}^{\infty}W' \left( u'(0^+) j + u'(0^-) i \right) \left[ u''(0^+) j^2 - u''(0^-) i^2 \right] \Bigg] 
 + o(\e^2) .
\end{split}
\end{equation}
Equalities \eqref{eq:ps1}, \eqref{eq:ps2}, \eqref{eq:ps3} and \eqref{eq:ps4} conclude the proof.
\end{proof}

\section{Sign of the jump term}\label{se:sign}

In this section we compare the conclusions of Theorems \ref{th:2ndorder} and \ref{th:2ndorderjump}.
The main difference in the assumptions is that in Theorem \ref{th:2ndorder} only smooth deformations are allowed, while in Theorem \ref{th:2ndorderjump} we allow deformations that are continuous and piecewise smooth.
In the corresponding Taylor expansion \eqref{eq:Eeu}, under the assumptions of Theorem \ref{th:2ndorderjump} this is reflected in the appearance of a jump-derivative term in the coefficient of order $\e$ and higher.
As explained in Section \ref{se:intro}, this term models the sharp-interface energy.
For physical reasons, we believe that this term should be positive, so that we need energy to create a (sharp) interface.
In this section, we analyse the sign of that term and show that, in many examples, it is indeed positive.

Let $W: \R \setminus \{0\} \to \R$ and $u: [-1,1] \to \R$ satisfy the same assumptions of Theorem \ref{th:2ndorderjump}.
Let $E^0$, $E^1$, $E^2$ be the coefficients defined in Theorem \ref{th:2ndorderjump}.
Let $\tilde{E}^0$, $\tilde{E}^1$, $\tilde{E}^2$ be, respectively, the coefficients \eqref{eq:E0}, \eqref{eq:E1}, \eqref{eq:E2}, once the substitutions
\[
 a=-1, \quad b=1, \quad b_1 = a_1 , \quad b_2 = a_2 , \quad c_1 = \frac{a_1}{2} - \frac{1}{4} , \quad c_2 = \frac{a_1^2}{2} - \frac{a_1}{2} + \frac{a_2}{2}+ \frac{1}{8}
\]
have been made.
Of course, those substitutions are motivated by Lemma \ref{le:[-1,1]}.
For each $i \in \{0,1,2\}$, call $J^i := E^i - \tilde{E}^i$.
Then $J^0 = 0$,
\begin{equation*}
 J^1 = - \frac{1}{4} \sum_{j=2}^{\infty} (j - 1) W(u'(0^-)j)
  - \frac{1}{4} \sum_{j=2}^{\infty} (j - 1) W(u'(0^+)j)
 + \frac{1}{2} \sum_{i,j= 1}^{\infty} W \left( u'(0^+) j + u'(0^-) i \right)  ,
\end{equation*}
\begin{align*}
 J^2 = & \frac{1}{2} \sum_{j=2}^{\infty} \left( \frac{1}{4} - \frac{a_1}{2} \right) (j-1) \left[ W(u'(0^-)j) + W(u'(0^+)j) \right]
 \\ & + \frac{1}{2} \sum_{j=2}^{\infty} 
        \left( \frac{1}{12}j - \frac{1}{4} j^2 + \frac{1}{6}j^3 \right) \left[ W'(u'(0^-)j) u''(0^-) - W'(u'(0^+)j) u''(0^+) \right]
 \\ & + \left( \frac{a_1}{2} - \frac{1}{4} \right) \sum_{i,j= 1}^{\infty} W \left( u'(0^+) j + u'(0^-) i \right) 
 \\ &  + \frac{1}{4} \sum_{i,j= 1}^{\infty} \left[ u''(0^+) j^2 - u''(0^-) i^2 \right] W' \left( u'(0^+) j + u'(0^-) i \right) .
\end{align*}
The equality $J^0 =0$ expresses the fact that jumps in the derivative do not affect the elastic energy, which corroborates the model \eqref{eq:I}, and is a known result in the $\G$-convergence approach of the problem (see, e.g., Braides \& Cicalese \cite{BC07}).
Thus, $J^1$ seems to represent (an approximation of a scaling of) the sharp-interface energy.

In order to ascertain the sign of $J^1$ we define $J: (0,\infty)^2 \to \R$ as
\begin{equation}\label{eq:J}
   J (a,b) := - \frac{1}{4} \sum_{j=2}^{\infty} (j - 1) W(a j)
  - \frac{1}{4} \sum_{j=2}^{\infty} (j - 1) W(b j)
 + \frac{1}{2} \sum_{i,j= 1}^{\infty} W \left( b j + a i \right) , \quad a,b>0 .
\end{equation}
We believe that nothing can be said in general about the sign of $J$, so we restrict ourselves to the analysis of the case when the jump of the derivative is small, i.e., when $a \simeq b$.

\begin{proposition}
Let $W \in \mc{C}^2 ((0,\infty))$ satisfy that there are $C, R>0$ and $\a>2$ such that
\[
 \left| W^{i)} (t) \right| \leq C t^{-\a-i} , \qquad t \geq R, \quad i \in \{0,1,2\} .
\]
Define $J: (0,\infty)^2 \to \R$ as \eqref{eq:J}, and $A: (0,\infty) \to \R$ as
\begin{equation}\label{eq:A}
  A(a) := \frac{1}{12} \sum_{j=2}^{\infty} (j-j^3) W''(a j) , \qquad a>0 .
\end{equation}
Then there exists a neighbourhood $\mc{U}$ of $\{(a,a) \in (0,\infty)^2 : A(a)>0 \}$ such that $J(a,b)\geq 0$ for all $(a,b) \in \mc{U}$.
\end{proposition}
\begin{proof}
Some easy but tedious calculations show that, for all $a>0$,
\begin{equation*}
 J (a,a) = 0, \qquad D J (a,a) = (0,0) , \qquad D^2 J (a,a) = \begin{pmatrix} A(a) & -A(a) \\ -A(a) & A(a) \end{pmatrix} .
\end{equation*}
Elementary calculus then shows that the function $G: (0,\infty)^2 \to \R$ defined as 
\[
 G(a,b) := \left\{ \begin{array}{lll}
 \frac{2 J(a,b)}{(a-b)^2} & \mbox{if} & a,b>0 \ \mbox{and} \ a \neq b \\
 A(a) & \mbox{if} & a>0 
\end{array} \right.
\]
is continuous.
Hence there exists a neighbourhood $\mc{U}$ of $\{(a,a) \in (0,\infty)^2 : A(a)>0 \}$ such that $G(a,b)> 0$ and $J(a,b)\geq 0$ for all $(a,b) \in \mc{U}$.
\end{proof}

We finish this section with an example of an interesting potential for which the function \eqref{eq:A} can be computed.
For $\s>0$, let $W_{\s} : (0,\infty) \to \R$ be the Lennard-Jones potential defined by
\begin{equation}\label{eq:LJ}
  W_{\s} (t) := \left( \frac{\s}{t} \right)^{12} - \left( \frac{\s}{t} \right)^6, \qquad t >0 .
\end{equation}
Some calculations show that if
\[
 \frac{a}{\s} > \left( \frac{26 [\zeta(11) - \zeta(13)]}{7 [\zeta(5) - \zeta(7)]} \right)^{1/6} \simeq 0.603431
\]
then $A_{\s} (a)>0$, where $A_{\s}$ is defined by \eqref{eq:A}, but having replaced $W$ with $W_{\s}$.
Recall that the minimum of the Lennard-Jones potential is at $2^{1/6} \s \simeq 1.12246 \s$, and, thus, in particular, $A_{\s}$ is positive at that value.

\section{Several sharp interfaces, well separated from each other}\label{se:separated}

When the deformation $u$ presents several sharp interfaces (i.e., $u'$ is discontinuous at finitely many points), and the interfaces are well separated from each other (i.e., the points of discontinuity of $u'$ do not depend on $\e$), then we have an exact analogue of Theorem \ref{th:2ndorderjump}, and, in particular, in the Taylor expansion of the energy, there is no term accounting for the interaction between sharp interfaces.

\begin{proposition}\label{prop:several}
Let $W : \R \setminus \{0\} \to \R$ be a $\mc{C}^{\infty}$ function such that $W(x) = W(-x)$ for all $x \in \R \setminus \{0\}$, and satisfy that there exist $C,R >0$ and $\a>3$ such that \eqref{eq:assumptionsW}.
Let $n \in \N$ and $t_0 < \cdots < t_{n+1}$.
Define $a:=t_0$ and $b:=t_{n+1}$.
Let $c \in \R$ and define $\ell := c +\Z$.
Let $u: [a,b] \to \R$ be continuous, increasing and such that $u|_{[t_p,t_{p+1}]}$ is a $\mc{C}^{\infty}$ diffeomorphism for each $p \in \{0,\ldots,n\}$.
Take \eqref{eq:a1a2b1b2c1c2R} and a sequence $\e \to 0^+$ such that \eqref{eq:a1a2b1b2c1c2} and, for each $p \in \{0,\ldots,n\}$,
 \begin{equation*}
\begin{split}
 & \e (c + k_1^p) - t_p = a_1^p \e + a_2^p \e^2 + o(\e^2) , \qquad  t_{p+1} - \e (c + k_2^p) = b_1^p \e + b_2^p \e^2 + o(\e^2) , \\
 & \frac{1}{N^p} \frac{1}{\e} = \frac{1}{t_{p+1}-t_p} + c_1^p \e + c_2^p \e^2 + o(\e^2) ,
\end{split}
\end{equation*}
for some $a_1^p, a_2^p, b_1^p,  b_2^p, c_1^p, c_2^p \in \R$, and where $k_1^p, k_2^p, N^p$ are defined according to \eqref{eq:k1k2N}, by replacing $a$ with $t_p$ and $b$ with $t_{p+1}$.
Let $u_{\e}$ be the restriction of $u$ to $\e \ell \cap [a,b]$, and define \eqref{eq:Eeue}.
Then \eqref{eq:Eeu}, where \eqref{eq:E0},
\[
 E_1 := \sum_{p=0}^n d_p \int_{t_p}^{t_{p+1}} \sum_{j=1}^{\infty} W(u'(x)j) \dd x + F_1(u'(a),u'(t_1^-),u'(t_1^+),\ldots,u'(t_n^-),u'(t_n^+),u'(b)) ,
\]
\begin{align*}
 E^2 : = & - \frac{1}{24(b-a)} \int_a^b  \sum_{j=1}^{\infty} W''(u'(x)j) u''(x)^2 j^4 \dd x
 + \sum_{p=0}^n e_p \int_{t_p}^{t_{p+1}} \sum_{j=1}^{\infty} W(u'(x)j) \dd x 
 \\ & + F_2 \left( u'(a),u'(t_1^-),u'(t_1^+),\ldots,u'(b),u''(a),u''(t_1^-),u''(t_1^+),\ldots,u''(b) \right) ,
\end{align*}
for some $d_p \in \R$ depending on $c_1$, $c_1^p$, $t_{p+1} - t_p$, $b - a$ (for $p \in \{0,\ldots,n\}$), some $e_p \in \R$ depending on $c_1$, $c_1^p$, $c_2$, $c_2^p$, $t_{p+1} - t_p$, $b - a$ (for $p \in \{0,\ldots,n\}$), and some $F_1 \in \mc{C}^{\infty} ((0,\infty)^{2n+2})$ and $F_2 \in \mc{C}^{\infty} ((0,\infty)^{2n+2} \times \R^{2n+2})$ depending on $W$, $a_1^p$, $a_2^p$, $b_1^p$, $b_2^p$, $c_1^p$, $c_2^p$, $t_{p+1}-t_p$ (for $p\in \{0,\ldots,n\}$).
\end{proposition}
\begin{proof}
The proof is very similar to that of Theorem \ref{th:2ndorderjump} and will only be sketched.

For each $p \in \{0,\ldots,n\}$ and $\e>0$ define
\[
 E_{\e}^p (u) := \frac{1}{2 N^p} \sum_{i \neq j \in \ell \cap \frac{1}{\e}[t_p,t_{p+1}]} W \left( \frac{u(\e j) - u(\e i)}{\e} \right) .
\]
From Theorem \ref{th:2ndorder} we know that $E_{\e}^p (u) = (E^0)^p + \e (E^1)^p + \e^2 (E^2)^p + o(\e^2)$, where the expression of $(E^0)^p$, $(E^1)^p$, $(E^2)^p$ is given by \eqref{eq:E0}, \eqref{eq:E1}, \eqref{eq:E2}, respectively, but replacing $a$, $b$, $a_1$, $a_2$, $b_1$, $b_2$, $c_1$, $c_2$ with $t_p$, $t_{p+1}$, $a_1^p$, $a_2^p$, $b_1^p$, $b_2^p$, $c_1^p$, $c_2^p$, respectively.
We express
\begin{align*}
  \sum_{i \neq j \in \ell \cap \frac{1}{\e}[a,b]} \!
 = & \sum_{p=0}^n \ \sum_{i \neq j \in \ell \cap \frac{1}{\e}[t_p,t_{p+1}]} \! 
  + 2 \sum_{p=0}^{n-1} \ \sum_{i \in \ell \cap \frac{1}{\e}[t_p,t_{p+1}]} \ \sum_{j \in \ell \cap \frac{1}{\e}[t_{p+1},t_{p+2}] \setminus \{ i \}}
 \\ & + 2 \sum_{p=0}^{n-2} \ \sum_{q=p+2}^n \ \sum_{i \in \ell \cap \frac{1}{\e}[t_p,t_{p+1}]} \ \sum_{j \in \ell \cap \frac{1}{\e}[t_q,t_{q+1}]}
 \\ &- \sum_{p=0}^n \ \sum_{i \in \ell \cap \frac{1}{\e}[t_p,t_{p+1}]} \ \sum_{j \in \ell \cap \{t_1,\ldots,t_n\} \setminus \{ i \}} \!
  - \! \sum_{i \in \ell \cap \{t_1,\ldots,t_n\}} \ \sum_{j \in \ell \cap \frac{1}{\e}[a,b] \setminus \{ i \}} \! .
\end{align*}

Now
\begin{align*}
 \frac{1}{2N} \sum_{p=0}^n \ & \sum_{i \neq j \in \ell \cap \frac{1}{\e}[t_p,t_{p+1}]} W \left( \frac{u(\e j) - u(\e i)}{\e} \right) 
 = \frac{1}{N} \sum_{p=0}^n N^p E_{\e}^p (u)
 \\ & = \sum_{p=0}^n \frac{t_{p+1}-t_p}{b-a} \left( E^0 \right)^p + \e \sum_{p=0}^n \left( \frac{t_{p+1}-t_p}{b-a} \left( E^1 \right)^p + \a_p \left( E^0 \right)^p  \right) 
 \\ & \quad + \e^2 \sum_{p=0}^n \left( \frac{t_{p+1}-t_p}{b-a} \left( E^2 \right)^p + \a_p \left( E^1 \right)^p  + \b_p \left( E^0 \right)^p \right) ,
\end{align*}
where, for each $p \in \{ 0,\ldots,n \}$, the number $\a_p \in \R$ depends on $c_1$, $c_1^p$, $t_{p+1} - t_p$, $b - a$, and the number $\b_p \in \R$ depends on $c_1$, $c_1^p$, $c_2$, $c_2^p$, $t_{p+1} - t_p$, $b - a$.

Arguing as in the proof of Theorem \ref{th:2ndorderjump}, it is easy to see that
\begin{align*}
 & \frac{1}{2N} \left[ 2 \sum_{p=0}^{n-1} \ \sum_{i \in \ell \cap \frac{1}{\e}[t_p,t_{p+1}]} \ \sum_{j \in \ell \cap \frac{1}{\e}[t_{p+1},t_{p+2}] \setminus \{ i \}} - \sum_{p=0}^n \ \sum_{i \in \ell \cap \frac{1}{\e}[t_p,t_{p+1}]} \ \sum_{j \in \ell \cap \{t_1,\ldots,t_n\} \setminus \{ i \}} \! \right.
  \\ & \hspace*{2.5em} \left. - \! \sum_{i \in \ell \cap \{t_1,\ldots,t_n\}} \ \sum_{j \in \ell \cap \frac{1}{\e}[a,b] \setminus \{ i \}} \! \right] W \left( \frac{u(\e j) - u(\e i)}{\e} \right)
 \\ & = B_1 \left( u'(t_1^-),u'(t_1^+),\ldots,u'(t_n^-),u'(t_n^+) \right) \e
 \\ & \quad + B_2 \left( u'(t_1^-),u'(t_1^+),\ldots,u'(t_n^+),u''(t_1^-),u''(t_1^+),\ldots,u''(t_n^+) \right) \e^2 + o(\e^2) ,
\end{align*}
for some $B_1 \in \mc{C}^{\infty} ((0,\infty)^{2n})$ and $B_2 \in \mc{C}^{\infty} ((0,\infty)^{2n} \times \R^{2n})$ depending on $W$, $a_1^p$, $a_2^p$, $b_1^p$, $b_2^p$, $c_1^p$, $c_2^p$ for each $p\in \{0,\ldots,n\}$, and $t_{p+1}-t_p$ for each $p\in \{1,\ldots,n\}$.

Finally,
\begin{equation}\label{eq:nointeraction}
 \frac{1}{N} \sum_{p=0}^{n-2} \ \sum_{q=p+2}^n \ \sum_{i \in \ell \cap \frac{1}{\e}[t_p,t_{p+1}]} \ \sum_{j \in \ell \cap \frac{1}{\e}[t_q,t_{q+1}]} W \left( \frac{u(\e j) - u(\e i)}{\e} \right) = O( \e^{\a-1}) ,
\end{equation}
and, of course, $\e^{\a-1} = o(\e^2)$.
\end{proof}

The reason why in Proposition \ref{prop:several} only local terms appear in the expansion of the energy but not an interaction term between interfaces is that the sharp interfaces are separated at a macroscopic distance.
We will see in the next section that if the sharp interfaces are separated at a microscopic distance then a new term will appear in the Taylor expansion accounting for that interaction energy.

\section{Sharp interfaces separated at an atomic scale: repulsion term}\label{se:repulsion}

From Proposition \ref{prop:several}, and especially from \eqref{eq:nointeraction}, we conclude that, in the model studied in Section \ref{se:separated} (which is virtually the same as that of Section \ref{se:sharp}), sharp interfaces do not interact with each other.
This is due to the decay conditions on $W$ and to the fact that the sharp interfaces are separated from each other at a macroscopic distance.
In this section, we will see that, if the sharp interfaces are separated at a distance comparable to the atomic one, then a small variant in the model predicts an interaction energy between two consecutive sharp interfaces.

In this paragraph, we briefly explain the atomistic model of this section and how it differs from the model of Section \ref{se:separated}.
The assumptions on the potential $W$ and on the atomistic energy \eqref{eq:Eeue} are the same.
As for the continuum deformation, we assume that $u$ has two sharp interfaces, separated at a distance which is a multiple of the interatomic distance $\e$.
The main difference is that the atomistic deformation $u_{\e}$ does not follow $u$ in the region between the two sharp interfaces.

Now we explain the model in more detail.
We assume that the macroscopic deformation $u$ presents exactly two sharp interfaces, which are at a distance of a fixed multiple of the atomistic scale; to be precise, at a distance $m \e$, for some integer $m \geq 2$.
It is not clear how to define the deformation in the region enclosed by the two interfaces, nor the energy associated to it.
Here we assume that, to the left of the leftmost sharp interface and to the right of the rightmost sharp interface, the atomistic deformation $u_{\e}$ follows the continuum deformation $u$, which is everywhere continuous and of class $\mc{C}^{\infty}$ outside $0$, whereas in the region between the two sharp interfaces, the atomistic deformation follows a scaling of a given deformation $y$, which in the end will solve an optimal profile problem.
Thus, we assume that there exist an increasing homeomorphism $u : [-1,1] \to \R$ such that $u |_{[-1,0]}$ and $u |_{[0,1]}$ are $\mc{C}^{\infty}$ diffeomorphisms, and an increasing function $y : [0,1] \to \R$ such that the atomistic deformation $u_{\e} : [-1,1] \cap \e \Z \to \R$ is defined by
\begin{align*}
 & u_{\e} |_{([-1,0] \cup [m \e,1]) \cap \e \Z} = u |_{([-1,0] \cup [m \e,1]) \cap \e \Z} , \\
 & u_{\e} (x) = \frac{u (m \e) - u (0)}{y(1) - y(0)} y(\frac{x}{m \e}) + \frac{u(0) y(1) - u(m \e) y(0)}{y(1) - y(0)} , \qquad x \in (0,m \e) \cap \e \Z .
\end{align*}
Take such a $y$.
Then, for all $a>0$ and $b \in \R$, the function $a y + b$ gives rise to the same $u_{\e}$.
Therefore, we can assume, without loss of generality, that $y(0) = 0$ and $y(1) = 1$.
Moreover, $y$ need not be defined in the whole $[0,1]$ but only on $\{0,\frac{1}{m},\ldots,1\}$.
Thus,
\[
 u_{\e} (j \e) = [ u (m \e) - u (0) ] y(\frac{j}{m}) + u(0) , \qquad 1 \leq j \leq m-1 .
\]
The atomistic energy associated with the deformation $u_{\e}$ is still \eqref{eq:Eeue}.
Since the model differs from that of Section \ref{se:sharp} only in what happens in $(0,\e m)$, it makes sense to compute the difference of the energies between the two models.
This is done in the next result.

\begin{theorem}\label{th:K1}
Let $W : \R \setminus \{0\} \to \R$ be a $\mc{C}^{\infty}$ function such that $W(x) = W(-x)$ for all $x \in \R \setminus \{0\}$, and satisfy that there exist $C,R >0$ and $\a>3$ such that \eqref{eq:assumptionsW}.
Let $u : [-1,1] \to \R$ be continuous, increasing and satisfy that $u|_{[-1,0]}$ and $u|_{[0,1]}$ are $\mc{C}^{\infty}$ diffeomorphisms.
Let $m\geq 2$ be a natural number.
Let $y: \{ \frac{1}{m} , \ldots, \frac{m-1}{m} \} \to (0,1)$ be a strictly increasing function.
For each $\e>0$, define $u_{\e} : [-1,1] \cap \e \Z \to \R$ by
\begin{align*}
 & u_{\e} |_{([-1,0] \cup [m \e,1]) \cap \e \Z} = u |_{([-1,0] \cup [m \e,1]) \cap \e \Z} , \\
 & u_{\e} (j \e) = [ u (m \e) - u (0) ] y(\frac{j}{m}) + u(0) , \qquad 1 \leq j \leq m-1 ,
\end{align*}
and let
\begin{align*}
 E_{\e} (u_{\e}) := & \frac{1}{2 \Card [-1,1] \cap \e\Z} \sum_{i \neq j \in \Z \cap \frac{1}{\e} [-1,1]} W \left( \frac{u_{\e}(\e j) - u_{\e}(\e i)}{\e} \right) , \\
 E_{\e} (u) := & \frac{1}{2 \Card [-1,1] \cap \e\Z} \sum_{i \neq j \in \Z \cap \frac{1}{\e} [-1,1]} W \left( \frac{u(\e j) - u(\e i)}{\e} \right)
\end{align*}
Take \eqref{eq:a1a2b1b2c1c2R} and a sequence $\e \to 0^+$ such that \eqref{eq:a1a2b1b2c1c2}.
Then $E_{\e} (u_{\e}) - E_{\e} (u) = \e K_1 + \e^2 K_2+ o(\e^2)$, where
\begin{equation}\label{eq:K1}
\begin{split}
  K_1 := & \frac{1}{2} \sum_{i=0}^{\infty} \sum_{j=1}^{m - 1} \left[  W \left( u'(0^+) m y(\frac{j}{m}) + u'(0^-) i \right) - W(u'(0^+)j + u'(0^-)i) \right]
 \\ & + \frac{1}{2} \sum_{i=1}^{m-2} \sum_{j=i+1}^{m-1} W \left( u'(0^+) m [ y(\frac{j}{m}) - y(\frac{i}{m}) ] \right)
 \\ & + \frac{1}{2} \sum_{i=1}^{m-1} \sum_{j=m}^{\infty} W \left( u'(0^+) \left[ j - m y(\frac{i}{m}) \right] \right) - \frac{1}{2} (m-1) \sum_{j=1}^{\infty} W(u'(0^+) j) ,
\end{split}
\end{equation}
\begin{align*}
 K_2 := & \frac{1}{4} \sum_{i=0}^{\infty} \sum_{j=1}^{m - 1} \left[ W' \left( u'(0^+) m y(\frac{j}{m}) + u'(0^-) i \right) \left[ u''(0^+) m^2 y(\frac{j}{m})  - u''(0^-) i^2 \right] \right.
 \\ & \hspace*{4.8em} \left. - W'(u'(0^+)j + u'(0^-)i) [ u''(0^+) j^2 - u''(0^-) i^2 ] \right]
 \\ & + \frac{1}{4} u''(0^+) \sum_{i=1}^{m-2} \sum_{j=i+1}^{m-1} W' \left( u'(0^+) m [ y(\frac{j}{m}) - y(\frac{i}{m}) ] \right) m^2 [ y(\frac{j}{m}) - y(\frac{i}{m}) ]
 \\ & + \frac{1}{4} u''(0^+) \sum_{i=1}^{m-1} \sum_{j=m}^{\infty} W' \left( u'(0^+) \left[ j - m y(\frac{i}{m}) \right] \right) \left[ j^2 - m^2 y(\frac{i}{m}) \right]
 \\ & - \frac{1}{4} u''(0^+) (m-1) \sum_{j=1}^{\infty} W'(u'(0^+) j) j (m+j) + \left( a_1 - \frac{1}{2} \right) K_1 .
\end{align*}
\end{theorem}
\begin{proof}
As in Lemma \ref{le:[-1,1]}, for each $\e>0$ let $k_2$ be the maximum integer less than or equal to $1/\e$, and define $N:= \Card [-1,1] \cap \e\Z$.
Then $N= 2k_2 +1$ and
\begin{equation}\label{eq:EeueEeu}
\begin{split}
  & E_{\e} (u_{\e}) - E_{\e} (u) =
 \\  & \frac{1}{N} \left[ \sum_{i=-k_2}^{0} \sum_{j=1}^{m - 1} + \sum_{i=1}^{m-2} \sum_{j=i+1}^{m-1} + \sum_{i=1}^{m-1} \sum_{j=m}^{k_2} \right] \!
 \left[ W \left( \frac{u_{\e} (\e j) - u_{\e} (\e i)}{\e} \right)
 - W \left( \frac{u (\e j) - u (\e i)}{\e} \right) \right] \! .
\end{split}
\end{equation}

Take $-k_2 \leq i \leq 0$ and $1 \leq j \leq m-1$.
Then
\begin{multline*}
  \frac{u_{\e} (\e j) - u_{\e} (\e i)}{\e} = \frac{u (\e m) - u (0)}{\e} y(\frac{j}{m}) + \frac{u (0) - u (\e i)}{\e}
 \\ = u'(0^+) m y(\frac{j}{m}) - u'(0^-) i
  + \frac{\e}{2} \left[ u''(0^+) m^2 y(\frac{j}{m})  - u''(0^-) i^2 \right]
 + O(\e^2) (-i)^3
\end{multline*}
and
\[
 \frac{u(\e j) - u (\e i)}{\e} = u'(0^+) j - u'(0^-) i + \frac{\e}{2} [ u''(0^+) j^2 - u''(0^-) i^2 ] + O(\e^2) (-i)^3 .
\]
Therefore,
\begin{align*}
 W \left( \frac{u_{\e} (\e j) - u_{\e} (\e i)}{\e} \right) 
 = & W \left( u'(0^+) m y(\frac{j}{m}) - u'(0^-) i \right) \\
 & + \frac{\e}{2} W' \! \left( u'(0^+) m y(\frac{j}{m}) - u'(0^-) i \right) \! \left[ u''(0^+) m^2 y(\frac{j}{m})  - u''(0^-) i^2 \right] \\
 & + O(\e^2) \left( (-i)^{3-a_1} + (-i)^{4-a_2} \right) + O(\e^4) (-i)^{6-a_2}
\end{align*}
and
\begin{align*}
 W \left( \frac{u(\e j) - u (\e i)}{\e} \right) = & W(u'(0^+)j - u'(0^-)i)
 \\ & + \frac{\e}{2} W'(u'(0^+)j - u'(0^-)i) [ u''(0^+) j^2 - u''(0^-) i^2 ]
 \\ & + O(\e^2) \left( (-i)^{3-a_1} + (-i)^{4-a_2} \right) + O(\e^4) (-i)^{6-a_2} .
\end{align*}
Hence, using \eqref{eq:a1a2b1b2c1c2} and Lemma \ref{le:[-1,1]},
\begin{equation}\label{eq:EeueEeu1}
\begin{split}
 \frac{1}{N} & \sum_{i=-k_2}^0 \sum_{j=1}^{m - 1} \left[ W \left( \frac{u_{\e} (\e j) - u_{\e} (\e i)}{\e} \right) - W \left( \frac{u(\e j) - u (\e i)}{\e} \right) \right]
 \\ & = \frac{\e}{2} \sum_{i=0}^{\infty} \sum_{j=1}^{m - 1} \left[  W \left( u'(0^+) m y(\frac{j}{m}) + u'(0^-) i \right) - W(u'(0^+)j + u'(0^-)i) \right]
 \\ & \quad + \left( \frac{a_1}{2} - \frac{1}{4} \right) \e^2 \sum_{i=0}^{\infty} \sum_{j=1}^{m - 1} \left[  W \left( u'(0^+) m y(\frac{j}{m}) + u'(0^-) i \right) - W(u'(0^+)j + u'(0^-)i) \right]
 \\ & \quad + \frac{\e^2}{4} \sum_{i=0}^{\infty} \sum_{j=1}^{m - 1} \left[ W' \left( u'(0^+) m y(\frac{j}{m}) + u'(0^-) i \right) \left[ u''(0^+) m^2 y(\frac{j}{m})  - u''(0^-) i^2 \right] \right.
 \\ & \hspace*{7.5em} \left. - W'(u'(0^+)j + u'(0^-)i) [ u''(0^+) j^2 - u''(0^-) i^2 ] \right] + o(\e^2) .
\end{split}
\end{equation}

Now take $1\leq i \leq m-2$ and $i+1 \leq j \leq m-1$.
Then
\begin{align*}
  \frac{u_{\e} (\e j) - u_{\e} (\e i)}{\e} & = \frac{u(m \e) - u(0)}{\e} [ y(\frac{j}{m}) - y(\frac{i}{m}) ] 
 \\ & = u'(0^+) m [ y(\frac{j}{m}) - y(\frac{i}{m}) ] + \frac{\e}{2} u''(0^+) m^2 [ y(\frac{j}{m}) - y(\frac{i}{m}) ] + O(\e^2) 
\end{align*}
and
\[
 \frac{u(\e j) - u (\e i)}{\e} = u'(0^+) (j-i) + \frac{\e}{2} u''(0^+) (j^2-i^2) + O(\e^2) .
\]
Therefore,
\begin{multline*}
 W \left( \frac{u_{\e} (\e j) - u_{\e} (\e i)}{\e} \right)
 = W \left( u'(0^+) m [ y(\frac{j}{m}) - y(\frac{i}{m}) ] \right)
 \\ + \frac{\e}{2} W' \left( u'(0^+) m [ y(\frac{j}{m}) - y(\frac{i}{m}) ] \right) u''(0^+) m^2 [ y(\frac{j}{m}) - y(\frac{i}{m}) ] + O(\e^2)
\end{multline*}
and
\[
 W \left( \frac{u(\e j) - u (\e i)}{\e} \right) = W(u'(0^+) (j-i)) + \frac{\e}{2} W'(u'(0^+) (j-i)) u''(0^+) (j^2-i^2) + O(\e^2) .
\]
Hence
\begin{equation}\label{eq:EeueEeu2}
\begin{split}
\frac{1}{N} & \sum_{i=1}^{m-2} \sum_{j=i+1}^{m-1} \left[ W \left( \frac{u_{\e} (\e j) - u_{\e} (\e i)}{\e} \right) - W \left( \frac{u(\e j) - u (\e i)}{\e} \right) \right]
 \\ & = \frac{\e}{2} \sum_{i=1}^{m-2} \sum_{j=i+1}^{m-1} \left[ W \left( u'(0^+) m [ y(\frac{j}{m}) - y(\frac{i}{m}) ] \right) - W(u'(0^+) (j-i)) \right] 
 \\ & \quad + \left( \frac{a_1}{2} - \frac{1}{4} \right) \e^2 \sum_{i=1}^{m-2} \sum_{j=i+1}^{m-1} \left[ W \left( u'(0^+) m [ y(\frac{j}{m}) - y(\frac{i}{m}) ] \right) - W(u'(0^+) (j-i)) \right] 
 \\ & \quad + \frac{\e^2}{4} u''(0^+) \sum_{i=1}^{m-2} \sum_{j=i+1}^{m-1} \left[ W' \left( u'(0^+) m [ y(\frac{j}{m}) - y(\frac{i}{m}) ] \right) m^2 [ y(\frac{j}{m}) - y(\frac{i}{m}) ] \right.
 \\ & \hspace*{11.5em} \left. - W'(u'(0^+) (j-i)) (j^2-i^2) \right] + o(\e^2) . 
\end{split}
\end{equation}

Finally, take $1 \leq i \leq m-1$ and $m \leq j \leq k_2$.
Then
\begin{align*}
  \frac{u_{\e} (\e j) - u_{\e} (\e i)}{\e} & = - \frac{u (\e m) - u (0)}{\e} y(\frac{i}{m}) + \frac{u (\e j) - u (0)}{\e}
 \\ & = u'(0^+) \left[ j - m y(\frac{i}{m}) \right]
  + \frac{\e}{2} u''(0^+) \left[ j^2 - m^2 y(\frac{i}{m}) \right]
 + O(\e^2) j^3
\end{align*}
and
\[
 \frac{u (\e j) - u (\e i)}{\e} = u'(0^+) (j-i) + \frac{\e}{2} u''(0^+) (j^2 - i^2) + O(\e^2) j^3 .
\]
Hence
\begin{align*}
 W \left( \frac{u_{\e} (\e j) - u_{\e} (\e i)}{\e} \right) = & W \left( u'(0^+) \left[ j - m y(\frac{i}{m}) \right] \right) 
 \\ & + \frac{\e}{2} W' \! \left( \! u'(0^+) \! \left[ j - m y(\frac{i}{m}) \right] \right) u''(0^+) \! \left[ j^2 - m^2 y(\frac{i}{m}) \right]
 \\ & + O(\e^2) (j^{3-a_1} + j^{4-a_2}) + O(\e^4) j^{6-a_2}
\end{align*}
and
\begin{align*}
 W \left( \frac{u (\e j) - u (\e i)}{\e} \right) = & W(u'(0^+) (j-i)) + \frac{\e}{2} W'(u'(0^+) (j-i)) u''(0^+) (j^2-i^2)
 \\ & + O(\e^2) (j^{3-a_1} + j^{4-a_2}) + O(\e^4) j^{6-a_2} .
\end{align*}
Therefore,
\begin{equation}\label{eq:EeueEeu3}
\begin{split}
 \frac{1}{N} & \sum_{i=1}^{m-1} \sum_{j=m}^{k_2} \left[ W \left( \frac{u_{\e} (\e j) - u_{\e} (\e i)}{\e} \right)
 - W \left( \frac{u (\e j) - u (\e i)}{\e} \right) \right] 
 \\ & = \frac{\e}{2} \sum_{i=1}^{m-1} \sum_{j=m}^{\infty} \left[ W \left( u'(0^+) \left[ j - m y(\frac{i}{m}) \right] \right) -  W(u'(0^+) (j-i)) \right]
 \\ & \quad + \left( \frac{a_1}{2} - \frac{1}{4} \right) \e^2 \sum_{i=1}^{m-1} \sum_{j=m}^{\infty} \left[ W \left( u'(0^+) \left[ j - m y(\frac{i}{m}) \right] \right) -  W(u'(0^+) (j-i)) \right]
 \\ & \quad + \frac{\e^2}{4} u''(0^+) \sum_{i=1}^{m-1} \sum_{j=m}^{\infty} \left[ W' \left( u'(0^+) \left[ j - m y(\frac{i}{m}) \right] \right) \left[ j^2 - m^2 y(\frac{i}{m}) \right] \right. 
 \\ & \hspace*{10.7em} \left. - W'(u'(0^+) (j-i)) (j^2-i^2) \right] + o(\e^2) .
\end{split}
\end{equation}
Equations \eqref{eq:EeueEeu}, \eqref{eq:EeueEeu1}, \eqref{eq:EeueEeu2} and \eqref{eq:EeueEeu3} conclude the proof.
\end{proof}

\section{The optimal profile problem}\label{se:optimal}

This section analyses the terms $K_1$ and $K_2$ of Theorem \ref{th:K1}.
As we explained in Section \ref{se:repulsion}, it is not clear how to define the atomistic deformation in the region between two sharp interfaces.
In Theorem \ref{th:K1}, we assumed that, in that region, the atomistic deformation $u_{\e}$ followed a scaling of a given discrete deformation $y : \{\frac{1}{m}, \ldots, \frac{m-1}{m} \} \to (0,1)$, but the actual values of $y$ were left unspecified.
We believe that $y$ should be such that the difference of energy $E_{\e} (u_{\e}) - E_{\e} (u)$ is minimum.
As this is a difficult problem, we approximate $E_{\e} (u_{\e}) - E_{\e} (u)$ by its Taylor expansion, and, thanks to Theorem \ref{th:K1}, we choose $y$ to be a minimiser of $K_1$.
For every $m$, this is a finite-dimensional minimisation problem, and it will turn out that, in many cases, the optimal choice of $y$ is the identity map, which gives $0$ as the optimal value for $K_1$.

As explained in Section \ref{se:intro}, the repulsion term between two sharp interfaces should be decreasing with respect to the distance between them, and tend to infinity as the interface goes to zero.
Since we have here a discrete variable $m$ that runs over $\{2,3,\ldots\}$, the latter property makes no sense, but we still can expect that the repulsion term is decreasing with respect to $m$, at least for small values of $m$.
In the context of Theorem \ref{th:K1}, the repulsion term is defined as $E_{\e} (u_{\e}) - E_{\e} (u)$, but again we approximate it by its Taylor expansion $\e K_1 + \e^2 K_2$.
As explained in the previous paragraph, the term $K_1$ provides us with information about the optimal shape of $y$ (which in many cases turns out to be the identity), but does not give much information about the repulsion energy (because in many cases it is zero) or about $m$ (because in many cases the optimal value of $K_1$ is $0$, regardless of $m$).
Thus, we will study the term $K_2$ and it will turn out that, in a particular but important case, the optimal $m$ is $6$.

The rest of this section is devoted to making those ideas precise and giving some examples in which the terms $K_1$ and $K_2$ can be calculated.

Given a function $W : (0,\infty) \to \R$ and a natural number $m\geq 2$, we define
\begin{equation}\label{eq:Um}
 \mc{U}_m := \left\{ (x_1, \ldots, x_{m-1}) \in \R^{m-1} : 0 < x_1 < \cdots < x_{m-1} < 1 \right\}
\end{equation}
and $F_m : (0,\infty)^2 \times \mc{U}_m \to \R$ as
\begin{equation}\label{eq:Fm}
\begin{split}
  F_m & (a,b; x_1, \ldots, x_{m-1}) := \sum_{i=0}^{\infty} \sum_{j=1}^{m - 1} \left[ W \left( b m x_j + a i \right) -  W(b j + a i)\right]
 \\ &+ \sum_{i=1}^{m-2} \sum_{j=i+1}^{m-1} W \left( b m (x_j - x_i) \right)
 + \sum_{i=1}^{m-1} \sum_{j=m}^{\infty} W \left( b \left[ j - m x_i \right] \right) - (m-1) \sum_{j=1}^{\infty} W(b j) ,
\end{split}
\end{equation}
for each $a,b>0$ and $(x_1, \ldots, x_{m-1}) \in \mc{U}_m$ for which all the series of \eqref{eq:Fm} converge.
Of course, the reason of this definition is that, according to Theorem \ref{th:K1} and specifically \eqref{eq:K1}, and following the notation there,
\[
 K_1 = \frac{1}{2} F_m \left( u'(0^-),u'(0^+); y(\frac{1}{m}), \ldots, y(\frac{m-1}{m}) \right) .
\]
In the next lemma we study the minimisers of $F_m$.
As we will see, the point $q_m \in \R^{m-1}$ defined by 
\begin{equation}\label{eq:qm}
 q_m := \left( \frac{1}{m},\ldots,\frac{m-1}{m} \right) 
\end{equation}
will play an important role.

\begin{lemma}\label{le:D2}
Suppose that $W \in \mc{C}^2 ((0, \infty))$ satisfies
\[
 \lim_{t\to 0+} W(t) = \infty, \qquad \limsup_{t\to \infty} \ t^{\a} \max \left\{ |W(t)| , |W'(t)|, |W''(t)| \right\} < \infty ,
\]
for some $\a >1$.
Let $m\geq 2$ be a natural number.
Define \eqref{eq:Um}, \eqref{eq:qm} and $F_m : (0,\infty)^2 \times \mc{U}_m \to \R$ as \eqref{eq:Fm}.
Let $a, b >0$.
Then there exists a minimiser of $F_m(a,b;\cdot)$ in $\mc{U}_m$.
Moreover, $F_m \left( a,b; q_m \right) = 0$ and $D F_m \left( a,a; q_m \right) = 0$.
Finally,
\begin{equation*}
 \p_{k \ell} F_m ( a,a; q_m ) = \left\{ 
 \begin{array}{lllll}
 \displaystyle 2 a^2 m^2 \sum_{j=1}^{\infty} W''(a j) & \mbox{if} & k , \ell \in \{1,\ldots,m-1\} & \mbox{with} & k = \ell \\
 \displaystyle - a^2 m^2 W'' \left( a |k-\ell| \right) & \mbox{if} & k , \ell \in \{1,\ldots,m-1\} & \mbox{with} & k \neq \ell .
\end{array}
 \right.
\end{equation*}
\end{lemma}
\begin{proof}
The assumptions imply that $F_m$ is of class $\mc{C}^2$, and, for each $a,b>0$ we have $F_m (a,b;x) \to \infty$ as $x \to \p \mc{U}_m$ with $x \in \mc{U}_m$.
This implies the existence of minimisers of $F_m (a,b;\cdot)$ in $\mc{U}_m$.
The rest of the lemma follows from a direct calculation.
\end{proof}

The next step would be to compute the minimisers of $F_m (a,b;\cdot)$ and to ascertain whether or not the point $q_m$ is a (local or global) minimiser of $F_m (a,a;\cdot)$.
The answer to these questions depends on $W$, $a$ and $b$; all we can say is that $q_m$ is not, in general, a critical point of $F_{m}(a,b;\cdot)$ for $a,b>0$ with $a \neq b$.
Since we do not think that there is a general answer to these questions, from now on we will concentrate on a specific example.

Let $\s>0$ and let $W_{\s} : (0,\infty) \to \R$ be the Lennard-Jones potential defined in \eqref{eq:LJ}.
From now on, for each natural $m \geq 2$, let the function $F_m$ defined in \eqref{eq:Fm} refer to the potential $W_{\s}$.
Then, a direct computation shows that, for all $a>0$,
\[
 \sum_{j=1}^{\infty} W''_{\s} (aj) = \frac{\s^6 a^{-14} \pi^8}{467775} \left( 8 \s \pi^6 - 2079 a^6 \right) ,
\]
\[
 \det D^2 F_3 \left( a,a;\frac{1}{3},\frac{2}{3} \right) = 4 \s^{12} a^{-28} \left[ \frac{\pi^{16}}{467775^2} \left( 8 \s^6 \pi^6 - 2079 a^6 \right)^2 - 9 \left( 26 \s^6 - 7 a^6 \right)^2  \right] .
\]
Based on those formulas and in Sylvester's Criterion, we find that
\[
 D^2 F_2 \left( a,a; \frac{1}{2} \right) > 0 \quad \mbox{if and only if} \quad \frac{a}{\s} < \left( \frac{8}{2079} \right)^{1/6} \pi ,
\]
\[
 D^2 F_3 \left( a,a;\frac{1}{3},\frac{2}{3} \right) > 0 \quad \mbox{if and only if} \quad \frac{a}{\s} < \left( \frac{\frac{8 \pi^{14}}{467775} -78}{\frac{\pi^8}{225} -21} \right)^{1/6} .
\]
Here we are using the following notation: if $A$ is a symmetric matrix, by $A > 0$ we mean that $A$ is positive definite.
Thus, we have a necessary and a sufficient condition for $q_m$ to be a local minimiser of $F_m (a,a;\cdot)$, for each $m \in \{2,3\}$.
In fact, numerical experiments with the software Mathematica \cite{Mathematica} suggest that $D^2 F_m (a,a;q_m) > 0$ if and only if $\det D^2 F_m (a,a;q_m) > 0$, if and only if $a< a_m \s$, where the numerical value of $a_m$ is shown in Table \ref{ta:am}.

\begin{table}[htb]
\small
\begin{center}
\begin{tabular}{p{.22\textwidth}p{.71\textwidth}}
\vspace*{-5.2cm}
\begin{tabular}{l|l}
 $m$ & $a_m$ \\ \hline
 $2$ & $1.24362$ \\
 $3$ & $1.24280$ \\
 $4$ & $1.24226$ \\
 $5$ & $1.24192$ \\
 $6$ & $1.24169$ \\
 $7$ & $1.24153$ \\
 $8$ & $1.24142$ \\
 $9$ & $1.24133$ \\
 $10$ & $1.24127$ \\
 $11$ & $1.24122$ \\
 $13$ & $1.24115$ \\
 $15$ & $1.24111$ \\
 $17$ & $1.24107$ \\
 $19$ & $1.24105$
\end{tabular} &
\includegraphics{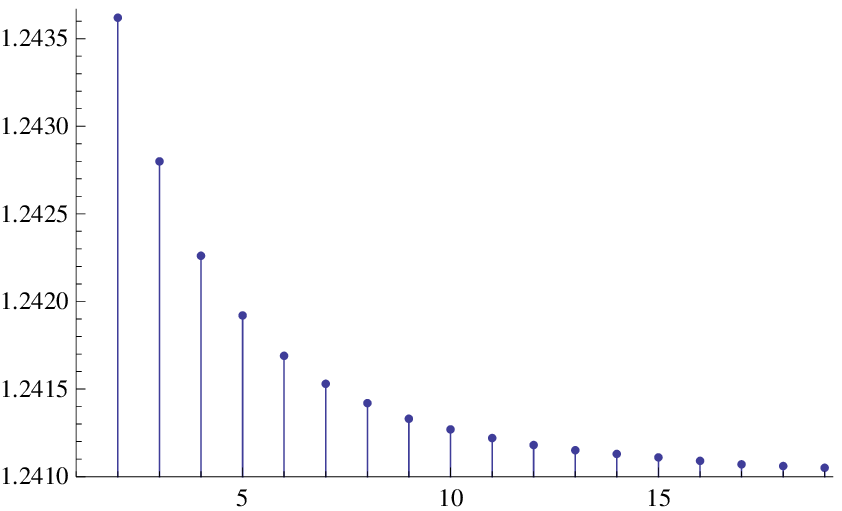}
\end{tabular}
\caption{Numerical values of $a_m$.}\label{ta:am}
\end{center}
\end{table}

Recall that there are two natural values of $a$, namely, the minimiser of $W_{\s}$, which is $2^{1/6} \s \simeq 1.12246 \s$, and the minimiser of the elastic energy (see Theorem \ref{th:2ndorder}), i.e., of $t \mapsto \sum_{j=1}^{\infty} W_{\s} (jt)$, which is $\left( \frac{1382}{675675} \right)^{1/6} \pi \s \simeq 1.1193 \s$.
Numerical experiments with Mathematica \cite{Mathematica} suggest that $q_m$ is in fact a global minimiser of $F_m (\tilde{a}\s,\tilde{a}\s;\cdot)$, where $\tilde{a} := \left( \frac{1382}{675675} \right)^{1/6} \pi$, for $m \in \{2,3,4,5\}$.

We sum up the findings up to now.
In the model described in Section \ref{se:repulsion}, the interaction energy term between the two sharp interfaces located at $0$ and $m \e$ is, by definition, $E_{\e}(u_{\e}) - E_{\e}(u)$, which depends on $u$, $y$, $\e$, $m$ (and, of course, $W$, but this is fixed beforehand).
Theorem \ref{th:K1} enables us to approximate $E_{\e}(u_{\e}) - E_{\e}(u)$ by $\e K_1$, so an approximation of a scaling of the interaction term is given by $K_1$, which depends on $u'(0^-)$, $u'(0^+)$, $y$, $m$.
Now we minimise in $y$, and hence the interaction term depends on $u'(0^-)$, $u'(0^+)$, $m$.
According to the analysis above, we believe that in many interesting cases, $q_m$ is a minimiser of $F_m (a,a;\cdot)$.
Assume that this is the case.
Then, by Lemma \ref{le:D2}, the first order term $K_1$ of the expansion of Theorem \ref{th:K1} is zero, regardless of $m$; hence, in order to have more information we study the second order term $K_2$.
As before, very little can be said about that term unless we assume a specific form for the potential $W$.
Motivated by the previous analysis, we define the function $G: (0,\infty)^2 \times \{2,3,\ldots\} \to \R$ as
\[
 G(a,\s,m) := \sum_{j=1}^{m-2} \left( -j^3 +2j^2 -j \right) W_{\s}'(a j) - (m-1) \sum_{j=m-1}^{\infty} (j-1) (2j-m) W_{\s}'(a j) ,
\]
for each $a,\s > 0$ and $m \in \{2,3,\ldots\}$.
It is easily checked that the expression $K_2$ of Theorem \ref{th:K1}, when the quantities $u'(0^-)$, $u'(0^+)$, $u''(0^-)$, $u''(0^+)$ have been replaced with $a$, $a$, $p$, $p$, respectively, and the functions $W$, $y$ have been replaced with $W_{\s}$, $\id$, becomes precisely $p G(a,\s,m) /4$.
For each $\s>0$, we choose $a_{\s} := \left( \frac{1382}{675675} \right)^{1/6} \pi \s$, and we compute the values of $G (a_{\s},\s,m)$.
It turns out that the quantity $\s G (a_{\s},\s,m)$ does not depend on $\s$, and their numerical values for several $m$ are displayed in Table \ref{ta:sG}, together with a graph of those values.
Again, the numerical values, as well as some symbolic calculations involving the Riemann zeta function, were obtained using Mathematica \cite{Mathematica}.

\begin{table}[htb]
\small
\begin{center}
\begin{tabular}{p{.22\textwidth}p{.71\textwidth}}
 \begin{tabular}{l|l}
 $m$ & $\s G (a_{\s},\s,m)$ \\ \hline
 $2$ & $-0.0570514$ \\
 $3$ & $-0.0657517$ \\
 $4$ & $-0.0470596$ \\
 $5$ & $-0.0453827$ \\
 $6$ & $-0.0452401$ \\
 $7$ & $-0.0452798$ \\
 $8$ & $-0.0453306$ \\
 $9$ & $-0.0453703$ \\
 $10$ & $-0.0453990$ \\
 $11$ & $-0.0454194$ \\
 $12$ & $-0.0454342$ \\
 $13$ & $-0.0454451$ \\
 $14$ & $-0.0454533$ \\
 $15$ & $-0.0454594$ \\
 $20$ & $-0.0454752$ \\
 $25$ & $-0.0454809$ \\
 $30$ & $-0.0454834$ \\
 $40$ & $-0.0454854$ \\
 $50$ & $-0.0454861$
\end{tabular}
& \includegraphics[scale=.94]{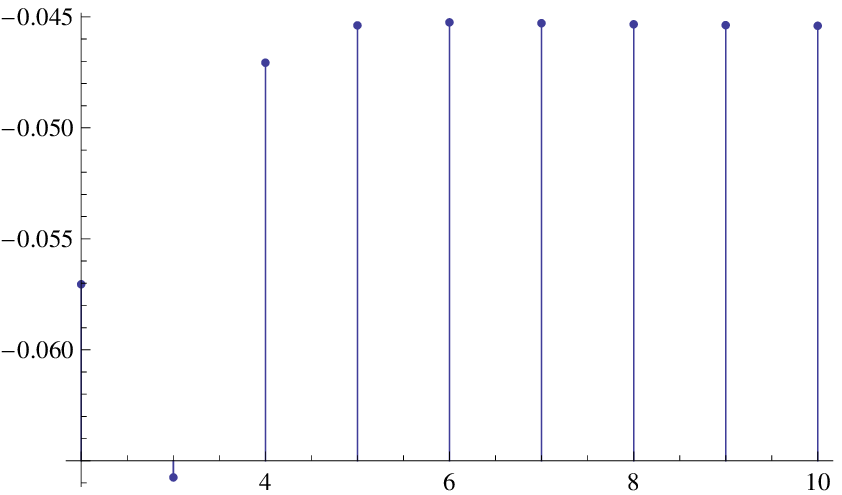}

\includegraphics[scale=.94]{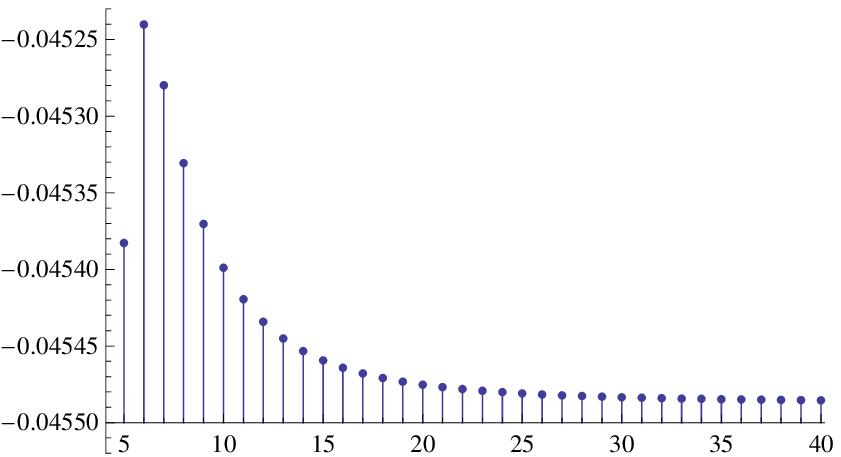}
\end{tabular}
\caption{Numerical values of $\s G (a_{\s},\s,m)$.\label{ta:sG}}
\end{center}
\end{table}

Our interpretation is the following.
For the Lennard-Jones potential, when $u'(0^-) \simeq u'(0^+) \simeq a/\s \simeq 1.1193$, then the optimal discrete configuration between two sharp interfaces separated at a distance $m \e$ is close to $q_m$.
If, in addition $u''(0^-) \simeq u''(0^+) > 0$ then the optimal $m$ is $6$, whereas if $u''(0^-) \simeq u''(0^+) < 0$ then the models predicts that no interfaces at all (i.e., $m=0$) is energetically better.
So this model predicts that the optimal length of the space between interfaces is $6$ times the atomistic distance, which coincides with the experiments of Baele, van Tendeloo and Amelinckx \cite{BVTA87}.

\subsection*{Acknowledgements}

It is a pleasure to thank John Ball for having proposed this problem to me and for many stimulating discussions.
The author has been supported by EPSRC (UK) and (partially) by Project CGL2006--00524/BOS (Ministry of Education and Science, Spain).

\end{document}